\documentstyle[aps,pra,eqsecnum,floats,amsfonts]{revtex}
\input epsf
\draft
\begin{document}
\title{MASSIVE FIELD-THEORY APPROACH TO SURFACE %
 CRITICAL BEHAVIOR IN THREE-DIMENSIONAL SYSTEMS}
\author{H.\ W.\ Diehl} 
\address{Fachbereich Physik, Universit\"at - Gesamthochschule  Essen,\\
D-45117 Essen, Federal Republic of Germany}
\author{M.\ Shpot\footnote{%
e-mail: shpot@icmp.lviv.ua}}
\address{Institute for
Condensed Matter Physics, 1 Svientsitskii str, 290011 Lviv, Ukraine}
\date{\today}
\maketitle
\begin{abstract}
The massive field-theory approach for studying critical behavior in
fixed space dimensions $d<4$ is extended to systems with surfaces.
This enables one to study surface critical behavior directly in
dimensions $d<4$ without having to resort to the $\epsilon$ expansion.
The approach is elaborated for the representative case of
the semi-infinite $|\bbox{\phi}|^4$ $n$-vector model
with a boundary term $\frac{1}{2}\,c_0\int_{\partial V}\bbox{\phi}^2$
in the action. To make the theory uv finite
in bulk dimensions $3\le d<4$, a renormalization of
the surface enhancement $c_0$ is required in addition to the
standard mass renormalization. Adequate normalization conditions for the renormalized theory are given. This theory involves two mass parameter:
the usual bulk `mass' (inverse correlation length) $m$, and
the renormalized surface enhancement $c$.
Thus the surface renormalization factors depend on
the renormalized coupling constant $u$ {\em and\/} the ratio $c/m$.
The special and ordinary surface transitions correspond to the
limits $m\to 0$ with $c/m\to 0$ and $c/m\to\infty$, respectively.
It is shown that the surface-enhancement renormalization turns into
an additive renormalization in the limit $c/m\to\infty$. The
renormalization factors and exponent functions with $c/m=0$ and $c/m=\infty$
that are needed to determine the surface critical exponents
of the special and ordinary transitions
are calculated to two-loop order. The associated
series expansions are analyzed by Pad\'e-Borel summation techniques. 
The resulting numerical estimates for the surface critical exponents are
in good agreement with recent Monte Carlo
simulations.
This also holds for the surface crossover
exponent $\Phi$, for which we obtain the values
$\Phi (n\!=\!0)\simeq 0.52$ and $\Phi (n\!=\!1)\simeq 0.54$
considerably lower than the previous $\epsilon$-expansion
estimates.
\end{abstract}

\section{Introduction}

Sparked by the emergence of renormalization group (RG) methods
at the beginning of the 1970s, the theory of bulk critical
phenomena has undergone a tremendous development in the past
25 years%
 \cite{WK74,DG76,ZJ}. Thanks to a
very fruitful interaction with field theory,
impressive progress has been achieved both in the theory
of bulk critical behavior and in field theory. While the latter
has provided a rich variety of powerful tools such as Feynman-graph
expansions and renormalized perturbation theory, on which analytical
RG approaches could be based, the former has offered
a wealth of challenging physical problems and served as a test
laboratory for the application of new field-theory techniques.

One popular line of approach that has been extensively used
with remarkable success are expansions about the upper critical
dimension $d^*$ ($=4$ for an ordinary bulk critical point)
\cite{WilsFish}.
The advantage of this technique is well known:
The computational effort required for calculations to low
orders in $\epsilon\equiv d^*-d$
is relatively modest, in particular, if the simplifying features
of such elegant schemes as dimensional regularization
and minimal subtraction of poles \cite{tHooft} are fully exploited.
As a consequence, the calculations can be ---
and have been \cite{glt84,kleinert} --- pushed to fairly high orders.

A major reason for this computational simplicity is that the
calculations can be performed directly
for the critical (massless) theory. However, there is a price
one must pay. The $\epsilon$ expansion involves a double
expansion in $\epsilon$ and $u$, the renormalized
coupling constant. In making this double expansion one by-passes
the problem that the perturbation series
of the critical theory in terms of the massless propagator
of the free theory is ill-defined for fixed $d<4$ because of infrared
singularities. In the dimensionally regularized theory, these
singularities manifest themselves as poles at rational values of
$\epsilon$ which accumulate at $d=d^*$ as the order of perturbation
theory increases \cite{Sympoles,bergere}. Thus the problem of summing these
infrared singularities arises. As stressed by Parisi \cite{Parisi},
without an additional hypothesis on the summation of these
singularities, any calculation based on the $\epsilon$ expansion
and the RG in this perturbative zero-mass scheme does not contain any
information about the critical behavior in a fixed dimension $d<d^*$.

In practice, the $\epsilon$ expansion often works amazingly well
for critical exponents, even if truncated at order $\epsilon^2$ and
extrapolated to $d=3$ in the most naive fashion
by setting $\epsilon=1$. However,
quantitatively accurate results require higher orders and sophisticated  
summation techniques \cite{glt84,kleinert,lgz85}.
The extrapolation problem usually is more
severe for other universal quantities such
as amplitude ratios%
\footnote{For a review of results on universal amplitude ratios, see Ref.\ \cite{ratios}.}
or scaling functions. One reason is that the results typically
involve (e.g., geometric) factors or functions
with an explicit dependence on $d$. Thus the question arises
whether and which of these $d$-dependent terms should be expanded
in $\epsilon$ or rather be kept in the extrapolation procedure.
As an empirical rule it has been advocated to choose the scale
of $u$ in such a fashion that a particular $d$-dependent geometrical
factor is absorbed \cite{Dohmempr}.
From a purely practical point of view, such recipes may well
be useful. But they are hardly satisfactory since they neither
have a firm theoretical basis we are aware of, nor
ensure that all ambiguities
of the extrapolation procedure are eliminated in a reliable fashion.

The field-theoretic RG approach based on the $\epsilon$
expansion has also been extended
\cite{DD80,DD81a,dd83,Syman,hwd,Die97} to,
and successfully used
in, the study of critical behavior of systems with {\it surfaces}%
\footnote{For a review of surface critical phenomena,
see Refs.\ \cite{hwd} and \cite{bdl};
 for an account of more recent results, see Ref.\ \cite{Die97}.}
\cite{hwd,Die97,bdl}. In the case of such systems an additional complication
may arise: even at low orders of the loop expansion, the perturbative
results may involve both geometric factors associated with
the $d$ dimensional bulk as well as others coming from
the $d-1$ dimensional boundaries. Hence
it may not even be clear how to apply the empirical rule
just mentioned.

From a fundamental point of view,
approaches that work directly in a {\it fixed dimension}
and therefore {\it avoid} the $\epsilon$ expansion
are clearly more attractive.
An important one of this kind is the
{\it massive field-theory approach
for fixed} $d<d^*$ \cite{Parisi,bgmn,bnm78,lgz80,bb85,bbmn,schd,parb,iz,ZJ}.
Its merits are well known:
Pushed to sufficiently high orders of perturbation theory
and combined with sophisticated series summation techniques,
it has produced values of bulk critical exponents
\cite{bnm78,lgz80} with an accuracy comparable to that of
the most precise ones
obtained so far by alternative methods
\cite{glt84,kleinert,lgz85,Pawley,liufish}, as well as
a set of amplitude ratios of barely inferior precision
\cite{bb85,bbmn,hald}.
The method has also been utilized,
albeit not to the same level of precision, to
determine the universal ratio of correlation-length amplitudes
for three-dimensional Ising systems
\cite{mu-h}, in the analysis of critical behavior in various
anisotropic and disordered systems \cite{shmn,mss,bsh},
partly even in general, non-integer
dimensions $2\le d<4$ \cite{jstph}, as well as
in studies of
three-dimensional $\phi^3$ theories describing the
percolation transition and the Yang-Lee edge singularity problem \cite{rgk82}.

In the present paper (a brief account of which has been given in  
Ref.\ \cite{Remprl}), we generalize the massive field-theory
approach for fixed $d<d^*$ to the study of critical
behavior in {\it semi-infinite systems}.
Such an extension is very desirable, both on account of
the general conceptual reasons explained above, and for
purely practical purposes. Recently extensive Monte Carlo calculations
\cite{ml88,lb90,vrf,rdww,hg94,rugew95}
have been performed for three-dimensional Ising models with free surfaces
and for the adsorption of polymers on walls%
\footnote{ For background and references on polymer adsorption on walls,
see Refs.\ \cite{ekb} and \cite{eekn}.}
\cite{ekb,eekn,ml88,hg94}. For most
surface critical exponents these yielded values in reasonable
agreement with the ones obtained by setting $\epsilon=1$ in their
$\epsilon$ expansion to order $\epsilon^2$ \cite{hwd}. For the
surface crossover exponent $\Phi$ \cite{hwd,bdl}, however,
the Monte Carlo estimates turned out to be
20--30\% lower. These discrepancies were
one of the motives for the present work.

Our analysis is based on
the semi-infinite $n$-vector model, which is the
appropriate prototype model for studying surface effects
on critical behavior \cite{hwd}.
In Sec.~\ref{Background} we briefly recall its definition
and provide the necessary background. In Sec.~\ref{NormCond} we give
normalization conditions for the massive field theory.
Sections \ref{SpTr}--\ref{SCESpTr} are devoted to the analysis of the {\it special\/} transition.
In Sec.\ \ref{SpTr} the general scheme of our approach is explained; then
the Callan-Symanzik equations are given and utilized to derive
the asymptotic scaling forms of the correlation functions
near the multicritical point describing the special transition.
After a brief discussion of some general features of perturbation
theory, our two-loop results for the RG functions are presented
in Sec.\ \ref{PT}.
These are utilized in Section VI to obtain numerical
estimates for the values of the surface critical exponents
of the special transition in three
dimensions by means of Pad\'e analyses and
Pad\'e-Borel summation techniques. The {\it ordinary\/} transition
is treated in Sec.~\ref{OrdTr}. Again, two-loop results are given and
exploited to obtain Pad\'e-Borel estimates of its surface critical exponents
for $d=3$.
Concluding remarks are reserved for Sec.~\ref{Concl}. 
Various calculational details are relegated to Appendices A and B.

\section{Background}\label{Background}

\subsection{The model}\label{Themodel}

Let $\bbox{\phi}=(\phi^a(\bbox{x}))$ be an $n$-vector field defined
on the half-space $V={\Bbb{R}}^d_+\equiv\{\bbox{x}=(\bbox{r},z)\in {\Bbb{R}}^d\mid
\bbox{r}\in {\Bbb{R}}^{d-1}, z\ge 0\}$ bounded by the plane $z=0$, which we
denote as $\partial V$. The semi-infinite $n$-vector model is
defined by the Euclidean action \cite{hwd,Die97}
\begin{eqnarray}\label{Ham}
{\cal H} [\bbox{\phi} ]&=&\int_V \left(\frac{1}{2}\,
\left(\partial_\mu\bbox{\bbox{\phi}}\right)^2
 + \frac{1}{2}\,m_0^2\,\bbox{\phi}^2+ \frac{1}{4!}\,u_0\,|\bbox{\phi} |^4\right)
+\int_{\partial V}\left(\frac{1}{2}\,c_0 \bbox{\phi}^2\right)\;.
\label{eh} \end{eqnarray}
Here $m_0^2$, $u_0$, and $c_0$ are the bare mass, the bare
coupling constant, and the bare
surface enhancement%
\footnote{Upon mapping a semi-infinite (lattice) Ising model
with ferromagnetic nearest-neighbor interactions of strength
$K_1$ between surface spins and of strength $K$ elsewhere
one finds that $c_0$ decreases as $(K_1-K)/K$ increases \cite{hwd}.
For simplicity, we shall nevertheless use the term surface enhancement
for $c_0$, rather than reserving it for
$(-c_0)$ or $(-c_0+\text{const})$.}%
, respectively.

Adding bulk and surface source terms to the action, we introduce the
generating functional
\begin{equation}\label{calZdef}
{\cal Z}[\bbox{J},\bbox{J_1};K,K_1]=
\int{\cal D}\bbox{\phi}\,\exp\left[-{\cal H}
+\int_V\left(\bbox{J}\!\cdot\bbox{\phi}+\frac{1}{2}
K\bbox{\phi}^2\right)
+\int_{\partial V}\left(\bbox{J_1}\!\cdot\bbox{\phi}+\frac{1}{2}  
K_1\bbox{\phi}^2\right)\right]
\end{equation}
and the correlation functions
\begin{eqnarray}\label{GNMII1}
\lefteqn{G^{(N,M;I,I_1)}\left(\bbox{x}_1,\ldots,\bbox{R}_{I_1}\right)}&&
\nonumber\\&=&
\left[\prod_{j=1}^N{\delta\over\delta J^{a_j}(\bbox{x}_j)}\right]
\left[\prod_{k=1}^M{\delta\over\delta J_1^{b_k}(\bbox{r}_k)}\right]
\left[\prod_{l=1}^I{\delta\over \delta K(\bbox{X}_l)}\right]
\left[\prod_{m=1}^{I_1}\left.{\delta\over\delta K_1(\bbox{R}_m)}\right]
\ln {\cal Z}\right|_{J=J_1=K=K_1=0}\,.
\end{eqnarray}
For the functions $G^{(N,M;0,0)}$ without $\bbox{\phi}^2$-insertions
on or off the surface we use the notation $G^{(N,M)}$. The tensorial indices
$\{a_j,b_k\}$ will be suppressed whenever no confusion is possible.
The ultraviolet (uv) singularities of the theory should be
assumed to be regularized by means of a large-momentum cutoff
$\Lambda$.

We shall also need the (bulk) analogs of these functions for the
$|\bbox{\phi}|^4$ theory in the infinite space, i.e., with $V={\Bbb{R}}^d$.
The easiest way to define these is the usual one where all
boundary terms in the action (\ref{Ham}) and the
generating functional (\ref{calZdef}) are dropped,
and periodic boundary conditions are chosen. We denote the so-defined
bulk analog of $G^{(N,0;I,0)}(\{\bbox{x}_j\};\{\bbox{X}_l\})$ as
$G_{\text{bulk}}^{(N;I)}(\{\bbox{x}_j\};\{\bbox{X}_l\})$ and introduce
their Fourier transforms $\tilde{G}_{\text{bulk}}^{(N;I)}$ through
\begin{eqnarray}\label{bulkGNI}
\lefteqn{G_{\text{bulk}}^{(N;I)}(\{\bbox{x}_j\};\{\bbox{X}_l\})}&&
\nonumber\\
&=&\int\limits_{\bbox{q}_1,\ldots,\bbox{Q}_I}
\tilde{G}_{\text{bulk}}^{(N;I)}(\{\bbox{q}_j\};\{\bbox{Q}_l\})\,
e^{i\left(\sum_j\bbox{q}_j\bbox{x}_j+\sum_l\bbox{Q}_l\bbox{X}_l\right)}\,
(2\pi )^d\delta\! \bigg(\sum_j\bbox{q}_j+\sum_l\bbox{Q}_l\bigg)\,,
\end{eqnarray}
where the integral  on the right-hand side indicates integrations  
$\int_{\bbox{q}}\equiv \int d^d(q/2\pi)$ over all $d$-dimensional
momenta $\bbox{q}_1,\ldots,\bbox{Q}_I$. For the associated standard bulk
vertex functions and their Fourier transforms we use
the notation $\Gamma_{\text{bulk}}^{(N;I)}$ and
$\tilde{\Gamma}_{\text{bulk}}^{(N;I)}$, respectively.

In the case of our half-space geometry, where
translational invariance is restricted
to translations {\it parallel} to the surface, it is
appropriate to perform Fourier transformations only with respect to
$(d-1)$-dimensional parallel coordinates. We denote the
$(d-1)$-dimensional parallel momenta associated with
the operators $\bbox{\phi}(\bbox{x}\notin \partial V)$ and  
$\bbox{\phi}_s\equiv\bbox{\phi}(\bbox{x}\in\partial V)$
by lower case $\bbox{p}$'s,
and those associated with the insertions
$\bbox{\phi}^2$ and $\bbox{\phi}_s^2$ by upper case $\bbox{P}$'s.
Parallel Fourier transforms are indicated by a hat; for example,
the pair correlation function in this $\bbox{p}z$ representation is
written as
$\hat{G}^{(2,0)}(p;z,z')$.

Infinitely far away from the surface all properties must attain their
bulk values. Hence the bulk functions $\hat{G}_{\text{bulk}}^{(N;I)}$
can be obtained from $\hat{G}^{(N,0;I,0)}$ by letting all $N+I$
perpendicular coordinates $z_j\to\infty$, keeping all
relative coordinates $z_{jk}\equiv z_i-z_k$ fixed:
\begin{equation}
\lim_{z_1,\ldots,z_{N+I}\to\infty }
\hat{G}^{(N,0;I,0)}(\{\bbox{p}\};\{z_{j}\})=
\hat{G}_{\text{bulk}}^{(N;I)}(\{\bbox{p}\};\{z_{jk}\})\;,
\end{equation}
where $\{\bbox{p}\}$ here stands for the set of all $N+I$
parallel momenta.

To proceed, it is necessary to recall a few well-known properties
of the model (\ref{Ham}) \cite{hwd}. Its phase diagram exhibits a
disordered phase (SD/BD), a surface-ordered, bulk-disordered
phase (SO/BD), and a surface-ordered,
bulk-ordered phase (SO/BO), provided $d$ exceeds the lower critical
dimension $d_{\text{SO/BD}}(n)$ for the appearance of a SO/BD phase.%
\footnote{This lower critical dimension  is given by $d_{\text{SO/BD}}(1)=2$
and $d_{\text{SO/BD}}(n)=3$ for the Ising case, $n=1$, and the $n$-vector
case with $n\ge 2$ and $O(n)$ symmetry, respectively. In the presence of
surface terms corresponding to an easy-axis  spin anisotropy, a SO/BD phase
is possible for $n\ge 2$ if $d>2$. This case, studied elsewhere \cite{DE84},
will not
be considered here. The case $d=3$ with $O(2)$ symmetry
of the Hamiltonian is special in that a surface phase with
{\em quasi-long-range order\/} can appear, a problem which will also not
be considered here. We shall also refrain from a discussion of
($d=2$)-dimensional $n$-vector models
with noninteger values of $n$ in the range $-2\le n\le 2$
(`loop models' \cite{Nienhuis}; see, e.g., \cite{Batchelor} and
its references). However, we shall estimate the surface critical exponents
for the $n\to 0$ case of  polymer adsorption \cite{ekb,eekn}, both
for the ordinary and special transition in $d=3$ dimensions.
}%

The boundaries between these phases
are the lines of surface, ordinary, and extraordinary transitions.
They meet at a multicritical point,
$(m_0^2,c_0)=(m_{0c}^2,c_0^{\text{sp}})$,
called special point and representing the special transition.
The ordinary and extraordinary transitions correspond
respectively to the
portions $c_0>c_0^{\text{sp}}$ and $c_0<c_0^{\text{sp}}$ of the
line of bulk criticality $m_0^2=m_{0c}^2$. The line of surface
transitions separates the SD/BD from the SO/BD phase.
At bulk criticality, we thus have three distinct transitions ---
the ordinary, special, and extraordinary transition. Of these only
the ordinary and special one can be reached from the disordered phase.
Since our present analysis is restricted to the disordered phase, only
these latter two types of transitions will be considered.

The restriction to the disordered phase simplifies the analysis
considerably. One does not have to deal with a nonvanishing, and
spatially varying, order-parameter profile
$\langle\bbox{\phi}(\bbox{x})\rangle$,
and the free propagator in the $\bbox{p}z$ representation takes
the relatively simple form
\begin{equation}\label{freeprop}
\hat{G}(\bbox{p};z,z')={1\over 2\kappa_0}\left[e^{-\kappa_0  
|z-z'|}-{c_0-\kappa_0\over
c_0+\kappa_0}\,e^{-\kappa_0 (z+z')}\right]
\end{equation}
with
\begin{equation}
\kappa_0=\sqrt{p^2+m_0^2}\;.
\end{equation}
The translation invariant first term on the right-hand side of
(\ref{freeprop}) is the free bulk propagator.

The perturbation series of the correlation
functions (\ref{GNMII1}) in terms of
the free propagator (\ref{freeprop}) can be regularized by setting  
$\hat{G}(\bbox{p};z,z')=0$ for $|\bbox{p}|>\Lambda$.  Whenever
we do not use dimensional regularization, the theory is understood
to be regularized in this fashion.

\subsection{Ultraviolet singularities for $d<4$}

Let us first discuss the uv singularities of the theory.
For bulk dimensions $d=4-\epsilon <4$ the theory is super-renormalizable.
Power counting shows \cite{hwd,Syman}
that the uv singularities
of the functions $\hat{G}^{(N,M)}$ can be absorbed through a mass shift
\begin{equation}
m_0^2=\hat m_0^2+\delta m_0^2
\end{equation}
and a surface-enhancement shift
\begin{equation}
c_0=\hat c_0+\delta c_0\;.
\end{equation}
In order that the $\hat G^{(N,M)}$
be finite for $2\le d<4$ when
expressed in terms of $\hat m_0^2$ and $\hat c_0$, the
contributions of order $u_0^\rho$ to these
shifts must behave as
\begin{equation}
\delta m_0^2\sim \Lambda^2 (u_0/\Lambda^\epsilon)^\rho
\end{equation}
and
\begin{equation}
\delta c_0 \sim \Lambda (u_0/\Lambda^\epsilon)^\rho
\end{equation}
in the limit $\Lambda\to\infty$. In contrast to $\delta m_0^2$, which
is known to be uv-divergent for $d\ge 2$, the shift $\delta c_0$
diverges only for $d\ge 3$.

\subsection{Poles of the dimensionally regularized theory}

As is well-known \cite{Sympoles,bergere,bb85}, if the theory
is regularized dimensionally, then the
uv singularities of the bulk correlation functions
manifest themselves as poles at $\epsilon=2/k$, with $k\in \Bbb{N}$.
These poles can be eliminated by means of an appropriate mass shift
$\delta m_0^2(\epsilon)$. We remind that  the
two-loop graph shown in Fig.\ 1
yields a contribution of the form
\begin{equation}
-u_0^{2}\,\frac{n+2}{18}\,m_0^{2-\epsilon}\,I_2(\epsilon)\;,\quad
I_2(\epsilon)\equiv
\int\limits_{\bbox{q},\bbox{q}'}{1\over\left(
q^2+1\right)\left(q'^2+1\right)
\left[\left(\bbox{q}+\bbox{q}'\right)^2+1\right]}\;\,,
\end{equation}
to
\begin{equation}
\tilde\Gamma_{\text{bulk}}^{(2)}(q=0)=\chi_{\text{bulk}}^{-1}\;,
\end{equation}
the inverse of the bulk susceptibility.
\begin{figure}[htb]
\centerline{\epsfbox{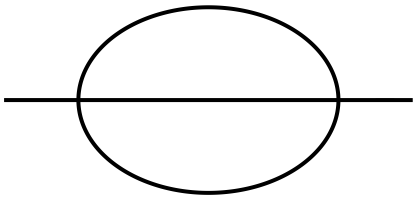}}
\caption{Two-loop Feynman diagram responsible for the poles
in the dimensionally regularized $\phi^4$ theory at $d=3$}
\label{fig1}
\end{figure}
The integral  has a simple pole
at $\epsilon=1$; i.\ e.,
$I_2(\epsilon)=R_2(\epsilon)/(\epsilon-1)$,
where $R_2(\epsilon)$ is regular at $\epsilon=1$ and whose value
$R_2(1)=1/32\pi^2$ can be calculated.
Removal of the pole is achieved by \cite{bb85}
\begin{equation}
\delta  
m_0^2(\epsilon)=u_0^{2/\epsilon}\,\frac{n+2}{18}\,
\frac{1}{32\pi^2\,(\epsilon-1)}\;.
\end{equation}

Expressed in terms of
$\hat m_0^2$ and $u_0$, the bare bulk functions
$\Gamma_{\text{bulk}}^{(N)}$ and $G_{\text{bulk}}^{(N)}$
are then finite at $d=3$. Yet, they also depend through logarithms
on $u_0$, and hence in a non-analytic fashion on it.

Not only does this non-analytic behavior carry over to the correlation
functions of our semi-infinite theory; owing to the appearance of
additional uv (surface) singularities, it shows up already at
one-loop order. To see this, consider the surface susceptibility
\begin{equation}
\chi_{11}(m_0,c_0)=
{\hat G}^{(0,2)}(p=0;m_0,c_0)\;.
\end{equation}
Its one-loop graph shown in Fig.\ 2 has a simple pole at $\epsilon=1$,
which can be removed by means of an appropriate choice of $\delta c_0$.

\begin{figure}[htb]
\centerline{\epsfbox{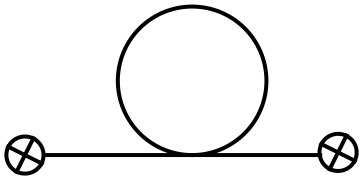}}
\caption{One-loop graph of the surface susceptibility
$\chi_{11}$ having a simple
pole at $d=3$. The crossed circles denote points on the surface.}
\label{fig2}
\end{figure}

We have
\begin{equation}
\chi_{11}(m_0,c_0)^{-1}=c_0+m_0+\frac{n+2}{6}\,
u_0\,m_0^{1-\epsilon}\,\Sigma_1\!\left(\epsilon,c_0/m_0\right )+{\cal  
O}(\text{2-loops})
\end{equation}
with 
\begin{eqnarray}
\Sigma_1(\epsilon,{\sf c})&=&\int_0^\infty  
dz\,e^{-2z}\int_{\bbox{p}}\frac{1}{2\sqrt{p^2+1}}\left[1-{{\sf  
c}-\sqrt{p^2+1}\over
{\sf c}+\sqrt{p^2+1}}\,e^{-2z\sqrt{p^2+1}}\right]
\nonumber\\&=&
J_1(\epsilon)+J_2(\epsilon,{\sf c})+J_3(\epsilon)\;,
\end{eqnarray}
where $\int_{\bbox{p}}\equiv (1/2\pi)^{d-1}\int d^{d-1}p$,
\begin{equation}
J_1(\epsilon)\equiv \int_{\bbox{p}}{1\over 4\sqrt{p^2+1}}  
=2^{-5-\epsilon}\pi^{-2+\epsilon/ 2}\,
\Gamma\!\left(-1+{\epsilon\over 2}\right)\,,
\end{equation}
\begin{equation}
J_2(\epsilon,{\sf c})\equiv -{\sf c\over 2}
\int_{\bbox{p}}\frac{1}{\sqrt{p^2+1}}\,
{1\over 1+\sqrt{p^2 +1}}\,{1\over {\sf c}+\sqrt{p^2 +1}}\;,
\end{equation}
and
\begin{equation}
J_3(\epsilon)\equiv \int_{\bbox{p}}{1\over 4\sqrt{p^2+1}}
\,{1\over 1+\sqrt{p^2 +1}}=\frac{1}{\epsilon -1}\,
2^{-3}\,\pi^{-1+\epsilon/2}\,
{\Gamma(\epsilon)\over\Gamma\!\left[(\epsilon+1)/2\right]}\;.
\end{equation}
The functions $J_1$ and $J_2$ are regular at $\epsilon=1$;
the above-mentioned pole results from $J_3$. Upon expansion of the
$\Gamma$-functions and computation of
\begin{equation}
J_2(1,{\sf c})={{\sf c}\ln 2-\ln ({\sf c}+1)\over 8\pi\,
(1-{\sf c})}\;,
\end{equation}
one arrives at the Laurent expansion
\begin{equation}
\Sigma_1(\epsilon,{\sf c})=\frac{1}{8\pi\,(\epsilon-1)}
+R_1({\sf c})+{\cal O}(\epsilon-1)\nonumber\\
\end{equation}
with
\begin{equation}
R_1({\sf c})=
{{\sf c}\ln 2-\ln ({\sf c}+1)\over 8\pi\, (1-{\sf c})}
-\frac{1}{32\pi}-\frac{C_E-\ln\pi}{16\pi}\;.
\end{equation}
Demanding that the pole be canceled by $\delta c_0$ gives
\begin{equation}
\delta c_0(\epsilon)=-u_0^{1/\epsilon}\,\frac{n+2}{48\pi}\,
{1\over \epsilon -1}\;.
\end{equation}
Expressed in terms of $\hat c_0$, $\hat m_0$, and $u_0$,
the bare susceptibility is finite at $d=3$. At  the level
of our one-loop calculation, one finds
\begin{equation}
\left.\chi_{11}^{-1}\right|_{\epsilon=1}
=\hat c_0+\hat m_0+\frac{n+2}{6}\,u_0\left[
R_1(\hat c_0/\hat m_0)-\frac{1}{8\pi}\ln {\hat m_0\over u_0}\right]
+{\cal O}(\text{2-loops})\;.
\end{equation}

The critical values $\hat m_{0c}^2=m_{0c}^2-\delta m_0^2$ and
$\hat c_0^{\text{sp}}=c_0^{\text{sp}}-\delta c_0$ of $\hat m_0^2$ and
$\hat c_0$ pertaining to the special point would have to be
determined from the conditions
\begin{equation}
\tilde\Gamma^{(2)}_{\text{bulk}}\!\left(q=0;m_{0c}^2\!
=\!\hat m_{0c}^2+\delta m_0^2\right)=0
\end{equation}
and
\begin{equation}
\chi_{11}^{-1}\!\left(m_{0c}\!=\!\sqrt{\hat m_{0c}^2+
\delta m_0^2},c_0^{\text{sp}}=\hat c_0^{\text{sp}}+\delta c_0\right)=0\;.
\end{equation}
The former is known to have the form \cite{Sympoles}
\begin{equation}
\hat m_{0c}^2=u_0^{2/\epsilon}\,\hat {\cal M}(\epsilon)\;.
\end{equation}
Similarly we have for the latter
\begin{equation}
\hat c_0^{\text{sp}}=u_0^{1/\epsilon}\,\hat{\cal C}(\epsilon)\;.
\end{equation}
The reason is that
 $u_0$ is the only dimensionful parameter
remaining at the special point
in the dimensionally regularized theory (with $\Lambda=\infty$
and $\epsilon>0$).
In view of the non-analytic dependence of   the susceptibilities on $u_0$,
it is clear that  Symanzik's observation
\cite{Sympoles} that $\hat {\cal M}(\epsilon)$
cannot be determined perturbatively carries over to $\hat{\cal C}(\epsilon)$.

In the next section we describe an appropriate extension of the
massive field theory approach that circumvents these difficulties.

\section{Normalization Conditions for the Massive Field Theory}
\label{NormCond}

Our aim is to study surface critical behavior
at a {\it bulk critical\/} point.
Therefore, a necessary property we ought to require
from our approach is that the bulk critical
behavior be treated appropriately.
A convenient way to achieve this
is to choose it in such a manner that it reduces
{\it for all bulk quantities\/} to a
well-established standard procedure. In our case this
will be the conventional one based on normalization
conditions (see, e.\ g., \cite{ZJ,parb,iz,blz,amit}).
Alternatively, one could choose an approach based on
minimal subtraction of poles of the massive theory,
as described for the bulk case by Schloms and Dohm \cite{schd}.

\subsection{Bulk normalization conditions}\label{BNC}

Starting from the bare bulk vertex functions  
$\Gamma_{\text{bulk}}^{(N,I)}(;m_0^2,,u_0)$, we perform a
mass shift
\begin{equation}
m_0^2=m^2+\delta m^2(\epsilon)
\end{equation}
and introduce  renormalization factors $Z_\phi (u)$,
$Z_{\phi^2}(u)$, $Z_u(u)$ (which are uv-finite for $d<4$)
as well as a renormalized
dimensionless coupling constant $u$
and  renormalized fields such that
\begin{equation}
\phi=\left[Z_\phi(u)\right]^{1/2}\phi_{\text{ren}}\;,\quad
\phi^2=\left[Z_{\phi^2}(u)\right]^{-1}  
\left[\phi^2\right]_{\text{ren}}\;,\quad u_0=Z_u(u)\,m^\epsilon \,u\;.
\end{equation}
The mass shift and the renormalization factors are fixed through the
standard normalization conditions
\begin{mathletters}
\begin{equation}\label{m2}
\left.\tilde\Gamma^{(2)}_{\text{bulk,ren}}\!\left(q;u,m\right)
\right|_{q
=0}=m^2\;,
\end{equation}
\begin{equation}\label{dgq}
\left.{\partial\over\partial q^2}\,
\tilde\Gamma^{(2)}_{\text{bulk,ren}}\!\left(q;u,m\right)
\right|_{q=0}=1\;,
\end{equation}
\begin{equation}\label{ng21}
\left.\tilde
\Gamma^{(2,1)}_{\text{bulk,ren}}\!
\left(\bbox{q},\bbox{Q};u,m\right)\right|_{\bbox{q}=\bbox{Q}=\bbox{0}}=1\;,
\end{equation}
\begin{equation}\label{normgamma4}
\left.\tilde
\Gamma^{(4)}_{\text{bulk,ren}}\!\left(\{\bbox{q}_i\};u,m\right)
\right|_{\{\bbox{q}_i=\bbox{0}\}}=
m^\epsilon\, u
\end{equation}
\end{mathletters}\noindent
for the renormalized vertex functions
\begin{equation}
\Gamma_{\text{bulk,ren}}^{(N,I)}
(\{\bbox{q},\bbox{Q}\};m,u)=\left[Z_\phi(u)\right]^{N/2}\,
\left[Z_{\phi^2}(u)\right]^I\,
\Gamma_{\text{bulk}}^{(N,I)}(\{\bbox{q},\bbox{Q}\};m_0^2,u_0)\;.
\end{equation}
Since  the mass shift is sufficient to absorb the uv singularities of the bare
functions $\Gamma_{\text{bulk}}^{(N,I)}$ at $d<4$, they become uv-finite when  
expressed in terms of $m$ and $u$ (or $m$ and $u_0$). The bulk renormalization
factors can be written as
\begin{mathletters}
\begin{equation}\label{Zphi-1}
\left[Z_\phi(u)\right]^{-1}=\left.{\partial\over\partial q^2}\,
\tilde\Gamma^{(2)}_{\text{bulk}}\!\left[q;m_0^2(m,u),u_0(m,u)\right]
\right|_{q=0}\;,
\end{equation}
\begin{equation}
\left[Z_{\phi^2}(u)Z_\phi(u)\right]^{-1}=\tilde
\Gamma^{(2,1)}_{\text{bulk}}\!\left[\{\bbox{0}\};m_0^2(m,u),u_0(m,u)\right],
\end{equation}
and
\begin{equation}
\left[Z_u(u)Z^2_\phi(u)\right]^{-1}=
\tilde\Gamma^{(4)}_{\text{bulk}}
\!\left[\{\bbox{0}\};m_0^2(m,u),u_0(m,u)\right]\!/u_0(m,u)\;.
\end{equation}
\end{mathletters}\noindent

\subsection{Surface normalization conditions}\label{ssnc}

Consider now the cumulants
$G^{(N,M)}$ and $G^{(N,M;I,I_1)}$
of our semi-infinite $\phi^4$ model.
As we have seen, a surface-enhancement shift $\delta c$
is required to
absorb uv singularities located on the surface.
Hence we write
\begin{equation}\label{deltac}
c_0=c+\delta c\;,
\end{equation}
where $c$ is a renormalized surface enhancement
whose precise definition we still have to give.

We also know that the surface operators
$\phi_s\equiv \phi|_{z=0}$ and $(\phi_s)^2$
should scale with scaling dimensions that are
different from those of  their bulk analogs $\phi(\bbox{x})$
and $[\phi(\bbox{x})]^2$ with $\bbox{x}\notin \partial V$.
This suggests the introduction of separate renormalization
factors for these surface operators, which we do via
the relations
\begin{equation}
\phi_s=\left[Z_\phi Z_1\right]^{1/2}
\left[\phi_s\right]_{\text{ren}}\;,\quad
(\phi_s)^2=\left[Z_{\phi_s^2}
\right]^{-1}\left[(\phi_s)^2\right]_{\text{ren}}
\end{equation}
between the bare and renormalized operators.
For the renormalized cumulants we thus have
\begin{equation}\label{gnmr}
G^{(N,M)}_{\text{ren}}(;m,u,c)=
Z_\phi^{-(N+M)/2}Z_1^{-M/2}G^{(N,M)}(;m_0,u_0,c_0)
\end{equation}
and
\begin{equation}
G^{(N,M;I,I_1)}_{\text{ren}}(;m,u,c)=
Z_\phi^{-(N+M)/2}Z_1^{-M/2}
\left[
Z_{\phi^2}
\right]
^I
\left[
Z_{\phi_s^2}
\right]
^{I_1}G^{(N,M;I,I_1)}(;m_0,u_0,c_0)\;.
\label{gnmir}\end{equation}

We wish to fix $\delta c$ and the new renormalization factors
$Z_1$ and $Z_{\phi_s^2}$ by appropriate normalization conditions.
To motivate our choice, let us recall the perturbation expansion
of the momentum-dependent surface
susceptibility $\chi_{11}(p)= \hat G^{(0,2)}(p)$
to lowest order:
\begin{equation}\label{sursuspertth}
\hat G^{(0,2)}(p;m_0,u_0,c_0)
={1\over c_0+\sqrt{p^2+m_0^2}}+{\cal O}(u_0)\;.
\end{equation}
We choose the normalization conditions such that the associated
renormalized susceptibility and its first derivatives with respect
to $p^2$  agree at $p=0$ with the corresponding
zero-loop expressions implied by  (\ref{sursuspertth}),
except for the replacements
$m_0\to m$ and $c_0\to c$. This gives
\begin{mathletters}
\begin{equation}\label{normc}
\left.\hat{G}_{\text{ren}}^{(0,2)}(p;m,u,c)
\right|_{p=0}=
{1\over m+c}
\end{equation}
and
\begin{equation}\label{normZ1}
{\partial\over\partial p^2}\,
\left.\hat{G}_{\text{ren}}^{(0,2)}(p;m,u,c)
\right|_{p=0}=-{1\over 2m(m+c)^2}\;.
\end{equation}
The following condition fixes the normalization of insertions
of the surface operator
$\case{1}/{2}\phi_s^2$, at zero external
momentum:
\begin{equation}\label{normZc}
\left. \hat{G}_{\text{ren}}^{(0,2;0,1)}(\bbox{p},\bbox{P};m,u,c)
\right|_{\bbox{p}=\bbox{P}=\bbox{0}}= (m+c)^{-2}\;.
\end{equation}
\end{mathletters}\noindent
This choice is motivated by the relation
\begin{equation}\label{gb0201}
\hat{G}^{(0,2;0,1)}(\{\bbox{0}\})=-{\partial\over\partial c_0}\,
\hat{G}^{(0,2)}(0)\;.
\end{equation}

Equation (\ref{normc}) defines the required surface-enhancement shift 
$\delta c$. Together with  (\ref{m2}), it ensures that the special
point is located at $m=c=0$. The
ordinary transition corresponds to the limit
$m\to 0$ at fixed $c>0$. In this limit  the renormalized surface susceptibility
$\chi_{11,\text{ren}} \to c^{-1}$. Hence the physical meaning of  $c$ is that
of the inverse of $\chi_{11,\text{ren}}$ at the transition.%
\footnote{Keeping $c$ (and $u$) fixed while changing $m$ requires,
of course, that the bare quantities $c_0$ (and $u_0$)
be varied with $m$. When exploiting the Callan-Symanzik equations
below, we shall as usual hold these bare quantities fixed while varying
$m$, so that the renormalized quantities $u$ and $c$ become $m$-dependent.}

Equations (\ref{normZ1}) and (\ref{normZc})
specify the renormalization factors $Z_1$ and $Z_{\phi_s^2}$,
respectively, in a similar manner as the bulk normalization conditions
(\ref{dgq}) and (\ref{ng21}) define $Z_\phi$ and $Z_{\phi^2}$.
The corresponding expressions are
\begin{mathletters}
\begin{equation}\label{zfz1}
Z_1Z_\phi=-2m(m+c)^2{\partial\over\partial p^2}\,
\left.\hat{G}^{(0,2)}\left[p;
m_0(m,u),u_0(m,u),c_0(c,m,u)\right]\,\right|_{p=0}
\end{equation}
and
\begin{equation}\label{zchwd}
Z_{\phi_s^2}^{-1}=-[Z_1Z_\phi]^{-1}(m+c)^2
{\partial\over\partial c_0}\,
\left.\hat{G}^{(0,2)}\left[0;m_0(m,u),u_0(m,u),c_0\right]
\,\right|_{c_0=c_0(c,m,u)}\;.
\end{equation}
\end{mathletters}

The above sets of normalization conditions
(\ref{m2})--(\ref{normgamma4}) and (\ref{normc})--(\ref{normZc}) define
$m_0^2$, $u_0$, $Z_\phi$, $Z_{\phi^2}$, $c_0$, $Z_1$, and $Z_{\phi_s^2}$
as functions of $m$, $u$, $c$, and $\Lambda$. All
$Z$ factors have  finite $\Lambda\to\infty$
limits in the $d<4$ case considered here.

For simplicity, we consider the $\Lambda=\infty$ limit in the
sequel. In our calculations described below
we actually took $\Lambda=\infty$ from the
outset, employing dimensional regularization.
In this limit the bulk $Z$ factors $Z_\phi$, $Z_{\phi^2}$, and $Z_u$
become functions of the single dimensionless variable $u$.
On the other hand,
the above choice of normalization conditions
(\ref{normc})--(\ref{normZc}) implies that
the surface $Z$ factors $Z_1$ and $Z_{\phi_s^2}$
depend on both $u$ and the dimensionless ratio $c/m$.

In a full investigation of the crossover
from the surface critical behavior characteristic
of the special transition (for $c/m\ll1$)
to that of the ordinary transition (for $c/m\gg 1$),
it would  be essential to carry along the
dependence on the variable $c/m$.
However, our main objective in the present work is
the calculation of the surface
critical exponents of the special and ordinary transitions.
To this end, a study of  the critical behavior
in the asymptotic limits $c/m\to 0$ and $c/m\to\infty$
is sufficient. As it turns out, there exist convenient
procedures (see \cite{Remprl} and below) 
which permit one to focus directly on these
limits, avoiding the need to keep track of the detailed dependence
on $c/m$.

\section{Special Transition}\label{SpTr}

Let us first consider the case of the {\it special\/} transition.
In order to reach the corresponding multicritical point,
we can safely set $c=0$.
This does not cause any problems
in the theory provided the surface-enhancement renormalization
has been performed. The desired critical behavior at the special
transition can then be obtained by studying the massless limit
of the resulting massive $c=0$  theory along lines analogous to those
usually followed in the bulk case. It follows that
the asymptotic critical behavior
at this transition is described
by the renormalized theory with the coupling constant
$u$ taken at  $u^*$, its value at the infrared-stable fixed point
(and $c$ set to zero).

\subsection{Normalization conditions at the multicritical point}\label{ssgc}

For $c=0$  the normalization conditions
(\ref{normc})--(\ref{normZc}) simplify. The $c=0$ analog of  (\ref{normc})
fixes the critical value $c_0^{\text{sp}}$ of $c_0$. Expressed in terms of
renormalized variables, it  takes the form $c_0^{\text{sp}}=mf_\epsilon(u)$ in
the dimensionally regularized  theory. Equations
(\ref{zfz1}) and (\ref{zchwd}) become
\begin{mathletters}
\begin{equation}
Z_1^{\text{sp}}(u)\,Z_\phi(u)=-2m^3{\partial\over\partial p^2}\,
\left.\hat{G}^{(0,2)}\big[p;
m_0(m,u),u_0(m,u),c_0^{\text{sp}}(m,u)\big]\,\right|_{p=0}\;,
\end{equation}
\begin{equation}
\left[Z_{\phi_s^2}^{\text{sp}}(u)\right]^{-1}=-m^2
\left[Z_1^{\text{sp}}(u)Z_\phi(u)\right]^{-1}{\partial\over\partial c_0}\,
\left.\hat{G}^{(0,2)}\big[
0;m_0(m,u),u_0(m,u),c_0\big]
\,\right|_{c_0=c_0^{\text{sp}}}\,,
\end{equation}
\end{mathletters}\noindent
specifying renormalization factors $Z_1^{\text{sp}}(u)\equiv Z_1(u,c/m=0)$ and
$Z_{\phi_s^2}^{\text{sp}}(u)\equiv Z_{\phi_s^2}(u,c/m=0) $
appropriate for the analysis of the special transition.
These renormalization factors
enter the relations (\ref{gnmr}) between the bare
and renormalized correlation functions $G^{(N,M)}$ for $c=0$,
\begin{equation}\label{gnmrsp}
G^{(N,M)}_{\text{ren,sp}}(;m,u)=
Z_\phi^{-(N+M)/2}(Z_1^{\text{sp}})^{-M/2}G^{(N,M)}(;m_0,u_0,c_0^{\text{sp}})\;,
\end{equation}
and the corresponding $c=0$ analogs of the relations (\ref{gnmir}) for $G^{(N,M;I,I_1)}$.

\subsection{Callan-Symanzik equations}\label{SCSE}

By varying $m$ at fixed $u_0$ and $c_0^{\text{sp}}$,
the Callan-Symanzik (CS) equations (cf.\ Refs. \cite{ZJ,parb,iz})
of the correlation functions $G^{(N,M)}_{\text{ren,sp}}$ can be
derived in a straightforward way. They read
\begin{mathletters}
\begin{equation}\label{CSE}
\left[m{\partial\over\partial m}
+\beta(u){\partial\over\partial u}
+{{N+M}\over 2}\,\eta_\phi(u)+{M\over 2}\,\eta_1^{\text{sp}}(u)\right]
G^{(N,M)}_{\text{ren,sp}}(;m,u)=\Delta G_{\text{ren}}\,,
\end{equation}
with
\begin{equation}\label{inhomo}
\Delta G_{\text{ren}} \equiv -\left[2-\eta_\phi(u)\right]m^2\,\int_V
d^dX\,G_{\text{ren,sp}}^{(N,M;1,0)}(;m,u)\;,
\end{equation}\end{mathletters}\noindent
where the integration is over the position $\bbox{X}$ of the inserted
$\phi^2$ operator.

The RG functions appearing here
are  the usual bulk functions
\begin{mathletters}
\begin{equation}\label{betadef}
\beta(u)=\left. m{\partial \over \partial m}\right|_0\,u
\end{equation}
and
\begin{equation}\label{etaphidef}
\eta_\phi(u)=\left. m{\partial \over \partial m}\right|_0\,\ln Z_\phi(u)
=\beta(u){d \ln Z_\phi(u)\over d u}\;,
\end{equation}
and the additional, surface-related function
\begin{equation}\label{eta1def}
\eta_1^{\text{sp}}(u)=\left. m{\partial\over\partial m}
\right|_0\,\ln Z_1^{\text{sp}}(u)
=\beta(u){d\, \ln Z_1^{\text{sp}}(u)\over d u}\;,
\end{equation}\end{mathletters}\noindent
where $|_0$
indicates that the derivatives are taken at fixed bare coupling
constant and surface enhancement (and cutoff $\Lambda$).

Just as in the bulk case, and as could be corroborated by means of a
short-distance expansion, the
right-hand side of (\ref{CSE}), $\Delta G_{\text{ren}}$,
should be negligible in the critical regime.
The resulting homogeneous equations
can be integrated in a standard fashion.

In order to identify the crossover exponent $\Phi$ we must also consider
deviations $\Delta c_0\equiv c_0-c_0^{\text{sp}}$ from the multicritical
point. We use the expansion
\begin{equation}\label{delcexp}
G^{(N,M)}(;m_0,u_0,c_0)=\sum_{I_1=0}^\infty\;{(-\Delta c_0)^{I_1}\over I_1!}\,
\underbrace{\int_{\partial V}\ldots\int_{\partial V}}_{I_1}
\,
G^{(N,M;0,I_1)}(;m_0,u_0,c_0^{\text{sp}})\;,
\end{equation}
where the integrations $\int_{\partial V}$ are over the positions of the $I_1$ inserted
$\phi^2_s$ operators.
No infrared problems arise here because the massive theory is used.

Expressing the right-hand side in terms of renormalized functions
and the renormalized variable
\begin{equation}\label{Deltacdef}
\Delta c\equiv\left[Z_{\phi_s^2}^{\text{sp}}(u)\right]^{-1}\Delta c_0\;
\end{equation}
gives 
\begin{equation}
Z_\phi^{-(N+M)/2}(Z_1^{\text{sp}})^{-M/2}G^{(N,M)}(;m_0,u_0,c_0)
=\sum_{I_1=0}^\infty\,{(-\Delta c)^{I_1}\over I_1!}\,
\underbrace{\int_{\partial V}\ldots\int_{\partial V}}_{I_1}\,
G^{(N,M;0,I_1)}_{\text{ren,sp}}(;m,u)
\;.
\end{equation}
Hence
\begin{equation}
G^{(N,M)}_{\text{ren}}(;m,u,\Delta c)
\equiv \left[Z_\phi(u)\right]^{-(N+M)/2}
\left[Z_1^{\text{sp}}(u)\right]^{-M/2}G^{(N,M)}(;m_0,u_0,c_0)
\end{equation}
are well-defined renormalized functions.\footnote{These should be %
distinguished from the previously defined $c$-dependent %
renormalized functions, which were related %
to the bare ones via $c$-dependent renormalization factors.}
Since they depend on the additional dimensionless variable
\begin{equation}\label{defsfc}
{\sf c} \equiv \Delta c/m\;,
\end{equation}
their RG equations are analogous to, but differ from, (\ref{CSE})
through the replacement
\begin{equation}
m{\partial\over\partial m}
+\beta(u){\partial\over\partial u}\quad\longrightarrow\quad
m{\partial\over\partial m}
+\beta(u){\partial\over\partial u}
-\left[1+\eta_{\sf c}^{\text{sp}}(u)\right]{\sf c}\,
{\partial\over\partial{\sf c}}\;,
\end{equation}
where
\begin{equation}\label{etacdef}
\eta_{\sf c}^{\text{sp}}(u)=\left. m{\partial \over \partial m}\right|_0\,
\ln Z_{\phi_s^2}^{\text{sp}}(u)
=\beta(u){d\over d u}\ln Z_{\phi_s^2}^{\text{sp}}(u)\;.
\end{equation}

\subsection{Scaling behavior near the multicritical point} 

The CS equations given in the preceding subsection
can be utilized in a familiar fashion to derive
the asymptotic scaling forms of the correlation functions
near the multicritical point. A detailed exposition of the
derivation of scale invariance and universality of bulk vertex functions
from the CS equations
may be found, for example,  in \cite{bb81} or elsewhere \cite{ZJ}.
Since in the present case a completely analogous line of reasoning
can be followed, we can be brief. In particular, we shall avoid
carrying along the various non-universal constants (metric factors
setting the scales of the relevant bulk and surface scaling
fields), as would be necessary for an explicit derivation of
four-scale-factor universality\cite{DGS85} within
the present massive RG framework.

We shall need the familiar dependence
\begin{equation}
m_0^2-m_{0c}^2\sim\tau
\end{equation}
of the bare mass on $\tau\equiv (T-T_c)/T_c$, valid for small
deviations from its critical value $m_{0c}^2$.
We also recall that $m$, which is nothing else than the
inverse of the (second-moment) bulk correlation length $\xi$,
behaves as
\begin{equation}
m\sim \left(m_0^2-m_{0c}^2\right)^\nu
\end{equation}
near criticality, with
\begin{equation}
\nu=[2+\eta^*_{\phi^2}]^{-1}\;,
\end{equation}
where $\eta^*_{\phi^2}$ denotes the value of the function
\begin{equation}
\eta_{\phi^2}(u)=
\left. m{\partial \over \partial m}\right|_0\,\ln Z_{\phi^2}(u)
=\beta(u)\,{d \ln Z_{\phi^2}(u)\over d u}\;,
\end{equation}
at the infrared-stable zero $u^*$ of $\beta(u)$.

Integration of  (\ref{betadef}), (\ref{etaphidef}), (\ref{eta1def}),
and (\ref{etacdef}) gives the asymptotic dependencies
\begin{equation}\label{Zphiasform}
Z_\phi\sim(u-u^*)^{\eta/\omega}\sim m^\eta\;,
\end{equation}
\begin{equation}
Z^{\text{sp}}_1\sim(u-u^*)^{\eta^{\text{sp}}_1(u^*)/\omega}
\sim m^{\eta^{\text{sp}}_1(u^*)}\;,
\end{equation}
and
\begin{equation}\label{Zcint}
Z^{\text{sp}}_{\sf c}\sim
(u-u^*)^{\eta_{\sf c}^{\text{sp}}(u^*)/\omega}\sim
m^{\eta^{\text{sp}}_{\sf c}(u^*)}
\end{equation}
for $u\to u^*$ or $m\to 0$. As usual, $\omega=\beta'(u^*)$ and
$\eta\equiv\eta_\phi(u^*)$. Equation (\ref{Zcint})
can be substituted into (\ref{Deltacdef}) and (\ref{defsfc}) to obtain
\begin{equation}
\Delta c\sim m^{-\eta_{\sf c}^{\text{sp}}(u^*)}\,\Delta c_0\;,\quad
{\sf c}\sim m^{-[1+\eta_{\sf c}^{\text{sp}}(u^*)]}\,\Delta c_0\;.
\end{equation}
From the latter result we read off the scaling variable
$\tau^{-\Phi}\,\Delta c_0$
with the crossover exponent
\begin{equation}\label{Phiscex}
\Phi=\nu [1+\eta^{\text{sp}}_{\sf c}(u^*)]\;.
\end{equation}

Using these results, one easily sees that
the CS equations of Sec.\ \ref{SCSE} yield the following asymptotic
scaling forms of the correlation functions near the multicritical
point:
\begin{equation}\label{scformsp}
G^{(N,M)}(\bbox{x},\bbox{r};m_0,u_0,c_0)\sim
m^{(N\beta+M\beta_1^{\text{sp}})/\nu}\,
\Psi^{(N,M)}\!\left(m\bbox{x},m\bbox{r},
m^{-\Phi/\nu}\Delta c_0\right)\;.
\end{equation}
Here $\beta$ and
$\beta_1^{\text{sp}}$ are standard bulk and surface exponents.
The latter is related to the usual surface correlation
exponent $\eta_\|$ via the scaling law
\begin{equation}
\beta_1=\frac{\nu}{2}\left(d-2+\eta_\|\right)\;,
\end{equation}
where $\eta_\|$ in the present case of the special transition
is given by
\begin{equation}\label{etapar}
\eta_\|^{\text{sp}}=\eta+\eta_1^{\text{sp}}(u^*)\;.
\end{equation}

\section{Perturbation Theory}\label{PT}

We now turn to the explicit calculation of the surface renormalization factors
$Z_1^{\text{sp}}$ and $Z_{\phi_s^2}^{\text{sp}}$ and of their associated
exponent functions $\eta_1^{\text{sp}}$ and $\eta_{\sf  c}^{\text{sp}}$.
In the one-loop approximation,
this will be carried out for general space dimensions $d<4$.
However, in our two-loop calculations we shall restrict ourselves to 
the case $d=3$.

\subsection{General Features}

The normalization conditions (\ref{normc}), (\ref{normZ1}), and (\ref{normZc})
we have chosen to fix  the surface counterterms all determine properties of the
renormalized surface susceptibility $\hat G^{(0,2)}_{\text{ren}}(p;m,u,c)$.
For calculational purposes it is more convenient to
express these conditions in terms of its bare inverse
$1/\hat G^{(0,2)}(p)$. From (\ref{normc}) we find
\begin{mathletters}\begin{equation}\label{cr}
Z_1Z_\phi \left[\hat{G}^{(0,2)}(0;m_0,u_0,c_0)\right]^{-1}=m+c\;.
\end{equation}
Utilizing this, (\ref{normZ1}) becomes
\begin{equation}
{\partial\over \partial p^2}
\left.\left[\hat G^{(0,2)}_{\text{ren}}(p;m,u,c)\right]^{-1}
\right|_{p=0} = {1\over 2m}\;,
\label{g0sec} \end{equation}\end{mathletters}\noindent
which shows that expression (\ref{zfz1})
for $Z_1Z_\phi$ can be cast in the equivalent
form
\begin{mathletters}\begin{equation}\label{zsp}
(Z_1Z_\phi)^{-1}=2m{\partial\over\partial p^2}\,
\left.\left[\hat{G}^{(0,2)}(p)\right]^{-1}\right|_{p=0}=
\lim_{p\to 0}{m\over p}{\partial\over \partial p}
\left[\hat{G}^{(0,2)}(p)\right]^{-1} \;.
\end{equation}
Likewise (\ref{zchwd})
can be rewritten as
\begin{equation}\label{zc}
Z_{\phi_s^2}^{-1}=Z_1Z_\phi{\partial\over\partial c_0}\,
\left[
\hat{G}^{(0,2)}(p=0)
\right]^{-1}.
\end{equation}\end{mathletters}\noindent

It is useful to decompose the above inverse surface susceptibility
into its free part, which is
$[\hat G(\bbox{p};0,0)]^{-1}=c_0+\kappa_0$
according to (\ref{freeprop}),
and a remainder due to perturbative corrections.
Thus we write
\begin{equation}\label{sigma0p}
\left[
\hat{G}^{(0,2)}(p)
\right]^{-1}
=c_0+\kappa_0-\hat\sigma_0(p)\;.
\end{equation}
To compute $\hat\sigma_0$, we start from the following
representation of the full propagator between two surface
points in terms of $\Sigma$, the usual `self-energy': 
\begin{equation}\label{Tmatrix}
G^{(0,2)}= 
\,_{s\,}\!|G|_s+\,_{s\,}\!|GTG|_s\;,\quad T=\Sigma \,(1-G\Sigma)^{-1}
\end{equation}
Here $_{s\,}\!|$ and 
$|_s$ indicate that the left and right points are located
on the surface, respectively.
A straightforward calculation yields
\begin{eqnarray}\label{sigmaexpans}
\hat\sigma_0(p)&=&
\frac{\hat{g}^{\text{T}}\hat T\hat g}%
{1+\,_{s\,}\!|\hat G|_s%
\,\hat{g}^{\text{T}}\hat T\hat g}
=\hat{g}^{\text{T}}\hat{\Sigma}\hat{g}
+\hat{g}^{\text{T}}\hat{\Sigma}\hat{G}\hat{\Sigma}\hat{g}-
\,_{s\,}\!|G|_s\left(\hat{g}^{\text{T}}\hat{\Sigma}\hat{g}\right)^2+
{\cal O}\Big(\hat\Sigma^3\Big)\;,
\end{eqnarray}
where $\hat g$ is a column vector whose components represent
the reduced propagator
\begin{equation}\label{redprop}
\hat g(p;z)\equiv(c_0+\kappa_0)\,\hat G(p;z,0)%
=e^{-\kappa_0z}\;,
\end{equation}
and $\hat{g}^{\text{T}}$ is its transpose.

The one-loop and two-loop contributions to
$\hat\sigma_0$ originating from
the first two terms on the right-hand side of
(\ref{sigmaexpans}) are depicted in
Fig.\ 3.  Denoting the one from the
graph labeled ``(i)'' by $C_i(p)$, 
we have
\begin{equation}\label{sigma0p2l}
\hat\sigma_0(p)=\sum_{i=1}^4C_i(p)
-{[C_1(p)]^2\over c_0+\kappa_0}+{\cal O}(u_0^3)\;,
\end{equation}
where the term $\propto C_1^2$ results from
the last one in  (\ref{sigmaexpans}).

\begin{figure}
\centerline{\epsfbox{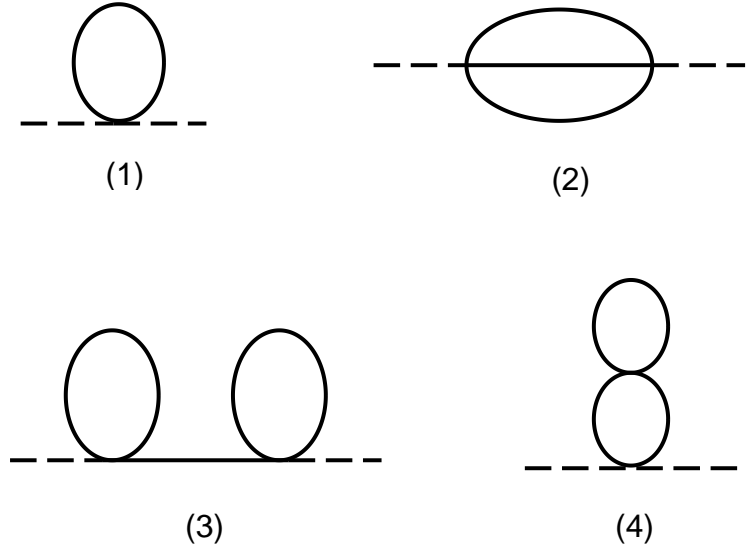}}
\caption{Feynman graphs to two-loop order of the nominator
$\hat{g}^{\text{T}}\hat T\hat g$ of  
the quantity ${\hat\sigma_0}(p)$ in
({\protect\ref{sigmaexpans}}).
Full lines denote the free propagator
({\protect\ref{freeprop}}), dashed ones
the reduced propagator
({\protect\ref{redprop})}.}
\end{figure}

\subsection{One-Loop Approximation}\label{sola}

We now specialize to the case $c=0$.
Upon using the above results together with
those presented in Appendix A, one can easily perform
a one-loop calculation of the renormalization functions
for general dimensions $d<4$. Following Ref.~\cite{bgmn},
let us introduce a rescaled renormalized coupling constant
$\tilde u$ through
\begin{equation}\label{utildedef}
u=b_n(d)\,\tilde u\;,\quad b_n(d)=\frac{6}{n+8}\,
\frac{(4\pi)^{d/2}}{\Gamma\big(\epsilon/2\big)}\;.
\end{equation}
The advantage of this choice is that
the expansion to order $\tilde u^2$ of the associated
beta function
$\tilde\beta=\beta\,\partial \tilde u/\partial u$
takes the simple form
\begin{equation}\label{betafunc}
\tilde\beta (\tilde u)=-\tilde u\,(1 -\tilde u)
+{\cal O}\Big(\tilde u^3\Big)\;.
\end{equation}
Our results for the two
surface renormalization factors of interest
then read
\begin{mathletters}
\begin{equation}
Z_1^{\text{sp}}
=1+{n+2\over n+8}\,{\tilde{u} \over 1+\epsilon}
+{\cal O}\Big(\tilde u^2\Big)
\label{z1ld}\end{equation}
and
\begin{equation}
Z_{\phi_s^2}^{\text{sp}}=1-{n+2\over n+8}\, 
\, {1 \over 1+\epsilon}\,
\left[1-2^{{1+\epsilon \over2}} {\,}_{2\!}F_1
\!\left(\!{3-\epsilon \over 2},{1+\epsilon \over 2};
{3+\epsilon \over 2};{1\over 2}\right) \right]\tilde{u}
+{\cal O}\Big(\tilde u^2\Big)\;,
\label{s39} 
\end{equation}\label{zzalld}
\end{mathletters}\noindent%
where
$_{2\!}F_1$ is the hypergeometric function \cite{as}. Substituting
these power series into (\ref{eta1def}) and (\ref{etacdef}) gives
\begin{mathletters}
\begin{equation}
\eta_1^{\text{sp}}(u)\equiv
\tilde\eta_1^{\text{sp}}(\tilde u)
=-{n+2\over n+8}\,
{\epsilon\over 1+\epsilon}\;\tilde u
+{\cal O}\Big(\tilde u^2\Big)\;,
\label{alldu}\end{equation}
\begin{equation}
\eta_{\sf c}^{\text{sp}}(u)\equiv
\tilde\eta_{\sf c}^{\text{sp}}(\tilde u)=
{n+2\over n+8}\,
{\epsilon\over 1+\epsilon}
\left[1-2^{{1+\epsilon \over2}} {\,}_{2\!}F_1\!
\left({3-\epsilon \over 2},{1+\epsilon \over 2};
{3+\epsilon \over 2};{1\over 2}\right) \right]
\tilde u+{\cal O}\Big(\tilde u^2\Big)\,.
\label{allduc}\end{equation}
\label{alldeta}\end{mathletters}\noindent

As a consistency check one can compute the pole
parts (PP) of the Laurent expansion of the above renormalization
factors at $\epsilon=0$. One finds
\begin{mathletters}
\begin{eqnarray}
\text{PP}_{\epsilon =0}\left[Z_1^{\text{sp}}-1\right]&=&
{n+2\over 3\epsilon} {u\over 16\pi^2}+{\cal O}\!\left(u^2\right)\\
&=&
 \text{PP}_{\epsilon =0}\!\left[Z_{\phi_s^2}^{\text{sp}}-1\right]
+{\cal O}\!\left(u^2\right)\,.
\end{eqnarray}
\end{mathletters}\noindent
As it should, this is in conformity with the one-loop terms
(of the two-loop results) of Ref.~\cite{dd83},
obtained by means of the usual
scheme of minimal subtraction of poles at $d=4$
for the massless theory in $4-\epsilon$
dimensions.\footnote{Since in Refs.~\cite{dd83} and \cite{hwd}
a factor $2^d\pi^{d/2}$ was absorbed
in the renormalized coupling constant, the
quantity $u/16\pi^2$ here
takes the place of the variable
$u$ of these references.}
Accordingly we also recover
the ${\cal O}(u)$ expressions
for the exponents functions
$\eta_1$ and $\eta_{\sf c}$ of  Ref.~\cite{dd83}
in the limit  $\epsilon\to 0^+$:
\begin{eqnarray}
\lim_{\epsilon\to 0^+}\eta_1^{\text{sp}}(u)&=&
-{n+2\over 3}\, {u\over 16\pi^2}+{\cal O}\!\left(u^2\right)
=\lim_{\epsilon\to 0^+}\eta_{\sf c}^{\text{sp}}(u)+{\cal O}\!\left(u^2\right)\,.
\end{eqnarray}

From (\ref{betafunc}) one reads off the
value $\hat u^*=1$ in this one-loop approximation.
Upon inserting this into  (\ref{alldeta}), we get
\begin{mathletters}\begin{eqnarray}\label{eta1-1loop}
\eta_1^{\text{sp}}(u^*)&=&
-{n+2\over n+8}\,{\epsilon\over 1+\epsilon}+{\cal O}(\text{2-loop})
=\eta_\|^{\text{sp}}+{\cal O}(\text{2-loop})
\label{eus}
\end{eqnarray}
and
\begin{equation}\label{etac-1loop}
\eta_{\sf c}^{\text{sp}}(u^*)={n+2\over n+8}\,
{\epsilon\over 1+\epsilon}\left[1-2^{{1+\epsilon \over2}}
{\,}_{2\!}F_1\!\left({3-\epsilon \over 2},{1+\epsilon \over 2};
{3+\epsilon \over 2};{1\over 2}\right) \right]+{\cal O}(\text{2-loop})\;.
\end{equation}
\end{mathletters}\noindent
The reader may check that the Taylor expansion
of these exponents to first order
in $\epsilon$ reproduces the known results
(see Refs.\ \cite{dd83,hwd} and references therein).

If we use (\ref{eta1-1loop}) and (\ref{etac-1loop})
to estimate their values for $d=3$, we find
\begin{mathletters}\label{eta1est-1l}\begin{eqnarray}
\eta_1^{\text{sp}}(n\!=\!0)&=&
\eta_\parallel^{\text{sp}}(n\!=\!0)
=-\frac{1}{8}\simeq -0.13\;,\\
\eta_1^{\text{sp}}(n\!=\!1)
&=&\eta_\parallel^{\text{sp}}(n\!=\!1)=-\frac{1}{6}\simeq -0.17\;,
\end{eqnarray}\end{mathletters}\noindent
and
\begin{mathletters}\label{etacest-1l}\begin{eqnarray}
\eta_{\sf c}^{\text{sp}}(n\!=\!0)&=&
\frac{1}{8}\left(1-4\ln2\right)\simeq -0.22\;,
\\
\eta_{\sf c}^{\text{sp}}(n\!=\!1)&=&
\frac{1}{6}\left(1-4\ln2\right)\simeq -0.30\;.
\end{eqnarray}\end{mathletters}\noindent
Amazingly, the estimates (\ref{eta1est-1l}) of this simple
calculation turn out to be 
among the best ones
for $\eta_\parallel^{\text{sp}}$ resulting from our much more
involved two-loop calculations
(see Tables \ref{Tsp1} and \ref{Tsp2}). On the other hand, our
best two-loop estimates for $\eta_{\sf c}^{\text{sp}}$
differ appreciably from
those listed in (\ref{etacest-1l}).

\subsection{Two-Loop Approximation}\label{stla}

At two-loop order we
restrict ourselves to the case $d=3$.
Details of the calculation may be found in
Appendix \ref{asp}.
Here we just quote the final results. They read
\begin{mathletters}\begin{equation}\label{endz1}
Z_1^{\text{sp}}Z_\phi=1+{n+2\over 2(n+8)}\,\tilde{u}
-{12\,(n+2)\over (n+8)^2} \left[
A-{n+2\over{12}}\,(1-\ln 2)\ln 2-{n+14\over48}\right]\tilde{u}^2
+{\cal O}\!\left(\tilde{u}^3\right)\;,
\end{equation}
\begin{eqnarray}\label{endzc}
Z_{\phi_s^2}^{\text{sp}}&=&1+{n+2\over  n+8}
\left(2\ln2-\frac{1}{2}\right)\tilde{u}
\nonumber\\&&\quad\mbox{}
+{12\,(n+2)\over (n+8)^2} \left[
A-B-{n\over2}\ln2+{n+2\over{2}}\ln^2 2
+{2n+1\over{12}}\right]\tilde{u}^2
+{\cal O}\!\left(\tilde{u}^3\right)\;,
\end{eqnarray}\end{mathletters}\noindent
\begin{mathletters}\begin{eqnarray}\label{etapsp}
\eta_\parallel^{\text{sp}}(\tilde{u})&=&
-{n+2\over 2(n+8)}\,\tilde{u}+{12\,(n+2)\over (n+8)^2} \biggl[ 2A
-{n+2\over6}\left(1-\ln2\right)\ln2
+{n-10\over48}\biggr]\tilde{u}^2
\nonumber\\&&\quad\mbox{}
+{\cal O}\!\left(\tilde{u}^3\right)\;,
\end{eqnarray}
and
\begin{eqnarray}
\eta_{\sf c}^{\text{sp}}(\tilde{u})&=&
-{n+2\over n+8}\left(2\ln2-{1\over2}\right)\tilde{u}
- {24\,(n+2)\over (n+8)^2} \biggl[ A-B
-\frac{n+1}{2}\,\ln2
\nonumber\\&&\quad\mbox{}
+{n+2\over3}\ln^2 2+{17n+22\over{96}}\biggr]\tilde{u}^2
+{\cal O}\!\left(\tilde{u}^3\right)\;,\label{etac}
\end{eqnarray}\label{alltwol}\end{mathletters}\noindent
where $A$ and $B$ are integrals originating from the two-loop
graph (2) of Fig.\ 3 whose values
\begin{equation}\label{Anumval}
A\simeq 0.202428
\end{equation}
and
\begin{equation}\label{Bnumval}
B\simeq 0.678061
\end{equation}
we have determined
by numerical means (cf.\ Appendix \ref{asp}).

\section{SURFACE CRITICAL EXPONENTS OF THE SPECIAL TRANSITION}%
\label{SCESpTr}

We shall now discuss how
the above perturbative results can be utilized
to estimate the surface critical exponents
of the special transition. Our starting point
are the series expansions of these
exponents in powers of $\tilde u$, which are implied by
(\ref{etapsp}) and (\ref{etac}). To generate these series,
we substitute the expansion (\ref{etapsp}) for $\eta^{\text{sp}}_\|$
into the following well-known
scaling-law expressions for surface exponents:
\begin{mathletters}
\begin{equation}
\Delta_1={\nu\over 2}\left(d-\eta_{\parallel}\right)\,,
\label{sc2}
\end{equation}
\begin{equation}
\eta_\perp ={\eta+\eta_{\parallel}\over 2}\;,
\label{sc3}
\end{equation}
\begin{equation}
\beta_1={\nu\over 2}\left(d-2+\eta_{\parallel}\right)\,,
\label{sc4} \end{equation}\begin{equation}
\gamma_{11}=\nu \left(1-\eta_{\parallel}\right)\,,
\label{sc5}
\end{equation}
\begin{equation}
\gamma_1=\nu\left(2-\eta_\perp \right)\,,
\label{sc6}
\end{equation}
\begin{equation}
\delta_1={\Delta\over \beta_1}= 
{d+2-\eta\over d-2+\eta_{\parallel}}\;,
\label{sc7}
\end{equation}
\begin{equation}
\delta_{11}={\Delta_1\over \beta_1}= 
{d-\eta_{\parallel}\over d-2+\eta_{\parallel}}\;.
\label{sc8}
\end{equation}
\label{allscal}\end{mathletters}\noindent
These scaling relations hold for the surface critical exponents
of the ordinary transition as well; therefore, we have omitted the
superscript ``sp''. We also need the expansions of the
bulk exponents $\nu$ and $\eta$. To the required order in $\tilde{u}^*$,
they read for the case $d=3$:
\begin{mathletters}\label{nuetaexp}
\begin{equation}\label{nuexp}
\nu(d\!=\!3,n)= {1\over 2}+
{n+2\over 4(n+8)}\: \tilde{u}^*+
{(n+2)(27n-38)\over 216 (n+8)^2}\:\left(\tilde{u}^*\right)^2
+{\cal O}\!\left[\left(\tilde{u}^*\right)^3\right]
\end{equation}
and
\begin{equation}\label{etaexp}
\eta(d\!=\!3,n)= {8(n+2)\over 27 (n+8)^2}\:\left(\tilde{u}^*\right)^2
+{\cal O}\!\left[\left(\tilde{u}^*\right)^3\right]\;.
\end{equation}
\end{mathletters}\noindent
We shall also consider the exponents
\begin{mathletters}
\label{aa}
\begin{equation}\label{alpha1sl}
\alpha^{\text{sp}}_1=\alpha+\nu-1+\Phi=1-
\nu\left[d-2-\eta_{\sf c}^{\text{sp}}(u^*)\right]
\end{equation}
and
\begin{equation}\label{alpha11sl}
\alpha^{\text{sp}}_{11}=\alpha+\nu-2+2\Phi=
-\nu\left[d-3-2\eta_{\sf c}^{\text{sp}}(u^*)\right]
\end{equation}
\end{mathletters}\noindent
of the layer and local specific heats $C_1(T)$ and  $C_{11}(T)$,
respectively \cite{hwd}. To obtain the expressions on
the extreme right-hand side, we have substituted
(\ref{Phiscex}) for $\Phi$ and used the
hyperscaling relation $\alpha=2-d\nu$.

For each one of these surface exponents we arrive at an
expansion of the form
\begin{equation}
f(\tilde{u}^*)=\sum_{k=0}^\infty f_k\left(\tilde{u}^*\right)^k
=f_0+f_1\,\tilde{u}^*+f_2\left(\tilde{u}^*\right)^2
+{\cal O}\!\left[\left(\tilde{u}^*\right)^3\right]\,.
\end{equation}
As is known from the much studied bulk case
(for background and references, see, e.g., \cite{ZJ}),
such series are {\it asymptotic\/}; they have
{\it zero radius of convergence\/}.
The reason for this is that the coefficients $f_k$ grow
proportional to $ k!$ as $k\to\infty$; more precisely,
their large-$k$ behavior typically can be written as
$f_k\approx {\cal C}\,k!\,k^{b-1}{\cal A}^{-k}$,
where the factor $k!$  basically reflects the enormous growth
of the number of diagrams contributing at a given order
of the loop expansion. We expect that these features will
carry over to the power series of surface quantities considered here.
The large-order behavior of their coefficients and the values
of the numbers ${\cal A}$, $b$, and ${\cal C}$ should be
obtainable by means of an appropriate extension of
the instanton calculus utilized in the case
of the $|\bbox{\phi}|^4$ bulk theory. Furthermore, 
in view of the rigorously established Borel summability of the
$d=3$ dimensional $|\bbox{\phi}|^4$
model \cite{Borelsumle},
we may be confident that these series
are Borel summable.

In order to obtain meaningful numerical estimates
from the above series expansions for surface critical
exponents, we must invoke appropriate and sufficiently
powerful summation techniques. The simplest
procedure is to construct the table
of Pad\'e approximants \cite{pade}.
This works well if successive elements $S_N$, $S_{N+1}$ of the
sequence of partial sums $S_N(\tilde{u}^*)\equiv
\sum_{k=0}^Nf_k(\tilde{u}^*)^k$ vary little at low orders of $N$.
A better and more sophisticated one is the Pad\'e-Borel
method used in Ref.\ \cite{bgmn}. At the order of perturbation
theory we are going to use it here, this involves the
analytic continuation of the Borel
transform
\begin{equation}
B_f(\tilde{u}^*)\equiv \sum_{k=0}^\infty \frac{f_k}{k!}
\left(\tilde{u}^*\right)^k
\end{equation}
by a $[1/1]$ Pad\'e approximant.

Our estimates given in Tables I--IV
were produced as follows. For each exponent $f$,
we rearranged the expansion as
$f/f_0\equiv M_f=1+(f_1/f_0)\tilde{u}^*+(f_2/f_0)(\tilde{u}^*)^2$
or
$f+(1-f_0)\equiv M_f=1+f_1\tilde{u}^*+f_2\,(\tilde{u}^*)^2$, depending
on whether $|f_0|>1$ or $|f_0|<1$, respectively. Then Pad\'e
approximants of the type indicated in Tables I--IV
were constructed for
the so-defined modified quantities $M_f$, and $[1/1]$ Pad\'e approximants
for their Borel transforms. For consistency reasons, these
approximants were evaluated using
the values of $\tilde{u}^*(d,n)$ one gets
from the Pad\'e-Borel resummed beta functions $\tilde\beta(\tilde u)$
at this two-loop order,
namely\footnote{The $n=0$ value $(\ref{fpv0})$
is given by the negative of
the value $v_2^*$ of the fixed point denoted U in Ref.~\cite{jug}.
}
\cite{bgmn,jug}
\begin{mathletters}
\begin{equation}\label{fpv0}
\tilde{u}^*(d\!=\!3,n\!=\!0)=1.632
\end{equation}
and
\begin{equation}
\tilde{u}^*(d\!=\!3,n\!=\!1)=1.597\,.
\end{equation}
\end{mathletters}\noindent
Finally, the resulting approximate values
of the $M_f$ were converted into estimates for the exponents
by inverting the above equations defining $M_f$ in terms of $f$.
Note that we used the hyperscaling relations with $d=3$. This is why
our zeroth approximations (gathered in the column $[0/0]$) do not always reproduce the Landau (or $\epsilon =0$) values
$0,\, 1,\, 0,\, \frac12,\, \frac12,\, 1,\,
3,\, 2$ and  $0,\, 0,\,-\frac12,\, \frac12$ of the exponents
$\eta_\|^{\text{sp}},\ldots,\,\delta_{11}^{\text{sp}}$
and $\eta_{\sf c}^*,\ldots,\Phi$
listed in the first columns of Tables I/II and III/IV, respectively.


\begin{table}[hbtp]\squeezetable
\caption[truc]{\label{Tsp1}
Surface critical
exponents of the special transition
involving  the RG function $\eta^{\text{sp}}_1$
for the case $n=0$ and $d=3$.}
\begin{tabular}{c||d|d|d|d|d|d|d|d|d|d|d}
 &${O_1/O_2}$&${O_{1i}/O_{2i}}$ &$[0/0]$ &
$[1/0]$ &$[0/1]$ &$[2/0]$ &$[0/2]$&
$[1/1]$ &$R $ &$R_i^{-1}$&
$f(\Delta_1 ,\nu ,\eta)$\\
\hline
$\eta_{\parallel}    $& $-$1.6&$-$2.4& 0.00& $-$0.204& $-$0.169
& $-$0.079& $-$0.107& $-$0.126& $-$0.134& $-$0.130& $-$0.133\\
$\Delta_1     $& $-$4.8&$-$2.4& 0.75& 0.954& 1.006
& 0.911& 0.886& 0.919& 0.921& 0.926& 0.921\\
$\eta_\perp   $& $-$1.4&$-$1.2& 0.00& $-$0.102& $-$0.114
& $-$0.027& $-$0.017& $-$0.059& $-$0.063& $-$0.064& $-$0.053\\
$\beta_1      $&  0.0&0.0& 0.25& 0.250& 0.250
& 0.264& 0.264&  -&  -&  -& 0.255\\
$\gamma_{11}  $& $-$3.6&$-$2.1& 0.50& 0.704& 0.756
& 0.647& 0.618& 0.660& 0.663& 0.668& 0.666\\
$\gamma_1     $& $-$4.5&$-$2.1& 1.00& 1.255& 1.342
& 1.199& 1.154& 1.209& 1.212& 1.220& 1.207\\
$\delta_1     $& $-$2.3&$-$1.6& 5.00& 6.020& 6.281
& 5.576& 5.397& 5.711& 5.739& 5.764& 5.734\\
$\delta_{11}  $& $-$2.4&$-$1.5& 3.00& 3.816& 4.121
& 3.481& 3.283& 3.578& 3.599& 3.628& 3.612\\
\end{tabular}
\end{table}

\begin{table}[hbtp]\squeezetable
\caption[truc]{\label{Tsp2}Surface critical
exponents of the special transition
involving  the RG function $\eta^{\text{sp}}_1$
for the case $n=1$ and $d=3$.}
\begin{tabular}{c||d|d|d|d|d|d|d|d|d|d|d}
& ${O_1/O_2}$&${O_{1i}/O_{2i}}$ & $[0/0]$ & $[1/0]$ & $[0/1]$ & $[2/0]$ & $[0/2]$ &
$[1/1]$ & $R $ & $R_i^{-1} $ & $f(\Delta_1 ,\nu ,\eta) $\\
\hline
$\eta_{\parallel}    $& $-$2.1&$-$4.8& 0.00& $-$0.266& $-$0.210
& $-$0.140& $-$0.174& $-$0.181& $-$0.189& $-$0.183& $-$0.165\\
$\Delta_1     $&$-$12.7&$-$2.9& 0.75& 1.016& 1.113
& 0.995& 0.961& 0.997& 0.997& 1.006& 0.997\\
$\eta_\perp   $& $-$1.7&$-$1.4& 0.00& $-$0.133& $-$0.154
& $-$0.056& $-$0.040& $-$0.084& $-$0.089& $-$0.091& $-$0.067\\
$\beta_1      $&  0.0&0.0& 0.25& 0.250& 0.250
& 0.261& 0.262&  -&  -&  -& 0.263\\
$\gamma_{11}  $& $-$8.2&$-$2.6& 0.50& 0.766& 0.863
& 0.734& 0.695& 0.737& 0.739& 0.747& 0.734\\
$\gamma_1     $&$-$11.0&$-$2.4& 1.00& 1.333& 1.499
& 1.302& 1.237& 1.305& 1.306& 1.320& 1.302\\
$\delta_1     $& $-$4.4&$-$2.0& 5.00& 6.331& 6.814
& 6.028& 5.779& 6.084& 6.101& 6.147& 5.951\\
$\delta_{11}  $& $-$4.8&$-$1.8& 3.00& 4.065& 4.650
& 3.845& 3.553& 3.882& 3.894& 3.947& 3.791
\end{tabular}
\end{table}

\begin{table}[hbtp]\squeezetable
\caption[truc]{\label{Tsp3}Surface critical exponents of the special transition
involving the RG function $\eta^{\text{sp}}_c$
for the case $n=0$ and $d=3$.}
\begin{tabular}{c||d|d|d|d|d|d|d|d|d|d|d}
& ${O_1/ O_2}$&${O_{1i}/ O_{2i}}$ & $[0/0]$ & $[1/0]$ & $[0/1]$ & $[2/0]$ & $[0/2]$ &
$[1/1]$ & $R $ & $R_i^{-1} $ & $f(\alpha_1,\nu ,\eta) $\\
\hline
$\eta_{\sf c}^*$&        $-$0.7&$-$0.9& 0.00& $-$0.362& $-$0.266& 0.183
& 0.055& $-$0.144& $-$0.168& $-$0.160& $-$0.119\\
$\alpha_1$&     $-$1.1&$-$1.7& 0.50& 0.217& 0.280& 0.467
& 0.399& 0.350& 0.336& 0.342& 0.342\\
$\alpha_{11}$&   $-$0.8&$-$1.1& 0.00& $-$0.362& $-$0.266& 0.109
& $-$0.021& $-$0.157& $-$0.180& $-$0.170& $-$0.140\\
$\Phi$&          $-$0.4&$-$0.4& 0.50& 0.421& 0.427& 0.642
& 0.657& 0.479& 0.474& 0.474& 0.518\\
\end{tabular}
\end{table}

\begin{table}[hbtp]\squeezetable
\caption[truc]{\label{Tsp4}Surface critical
exponents of the special transition
involving the RG function $\eta^{\text{sp}}_c$
for  the case $n=1$ and $d=3$.}
\begin{tabular}{c||d|d|d|d|d|d|d|d|d|d|d}
 & ${O_1/ O_2}$&${O_{1i}/ O_{2i}}$ & $[0/0]$ & $[1/0]$ & $[0/1]$ & $[2/0]$ & $[0/2]$ &
$[1/1]$ & $R $ & $R_i^{-1} $ & $f(\alpha_1,\nu ,\eta) $\\
\hline
$\eta_{\sf c}^*$&        $-$0.7&$-$1.1& 0.00& $-$0.472& $-$0.321& 0.168
& $-$0.052& $-$0.200& $-$0.230& $-$0.215& $-$0.144\\
$\alpha_1$&     $-$1.4&$-$2.9& 0.50& 0.131& 0.230& 0.393
& 0.304& 0.284& 0.268& 0.279& 0.279\\
$\alpha_{11}$&   $-$0.9&$-$1.6& 0.00& $-$0.472& $-$0.321& 0.042
& $-$0.153& $-$0.226& $-$0.253& $-$0.237& $-$0.182\\
$\Phi$&          $-$0.4&$-$0.4& 0.50& 0.397& 0.407& 0.649
& 0.661& 0.470& 0.463& 0.463& 0.539
\end{tabular}
\end{table}

The quantities $O_1$, $O_2$, $O_{1i}$, and $O_{2i}$
appearing in Tables I--IV
are defined through the expansions $M_f=1+O_1+O_2+\ldots
=[1+O_{1i}+O_{2i}+\ldots]^{-1}$ of the modified
quantities $M_f=f/f_0$
or $M_f=f+1-f_0$. Using the latter quantity 
to generate such truncated expansions in the case $|f_0|<1$
rather than simply factoring out $f_0$
yields better behaved `inverse expansions', i.e., series
for $M_f^{-1}$.

The bigger the absolute
values of the ratios $O_1/O_2$ and $O_{1i}/O_{2i}$ are,
the better is the quality of the resulting series for $M_f$ and
its reciprocal $1/M_f$, respectively.
All ratios $O_{1(i)}/O_{2(i)}$ 
have negative sign or else vanish.
Thus all series produced by these expansions
are alternating, and hence adapted to the
above-mentioned Pad\'e-Borel summation technique. (If
a series were not alternating, it would be unsuitable for
this method because the [1/1] approximant of its Borel transform
would have a pole on the positive real axis, i.\ e.,
inside the integration range \cite{bnm78}.) The estimates
obtained via Pad\'e-Borel resummation of the power series
for $M_f$ and $1/M_f$ are listed
in Tables I--IV as $R$ and $R_i^{-1}$, respectively.

In most cases the resulting power series in $\tilde u$
have 
second-order corrections $O_{2(i)}$
whose absolute values are smaller than those of
their first-order ones. Thus
the sequences of associated
partial sums
appear to be slowly convergent,
to the available low order.
Exceptions are some series
involving $\eta_{\sf c}^{\text{sp}}$, whose behavior is rather bad.
In the first group of exponents, related to $\eta^{\text{sp}}_1(u)$
and shown in Tables I and II, the 
most reliable estimates are obtained from the direct series
for the exponent $\Delta_1$, which appear to exhibit the best convergence
properties. These estimates are $\Delta_1=0.921$ for $n=0$
and $\Delta_1=0.997$ for $n=1$. Substituting these along with
the standard bulk values \cite{lgz80,ZJ} $\nu(n\!=\!0)=0.588$,
$\eta(n\!=\!0) =0.027$, $\nu(n\!=\!1)  =0.630$, and
$\eta(n\!=\!1)=0.031$ into the scaling laws (\ref{sc2})--(\ref{sc8}),
we have computed the remaining seven exponents of this group. 
The resulting values $f(\Delta_1 ,\nu ,\eta)$
are presented in the last columns of Tables
\ref{Tsp1} and \ref{Tsp2}.
By and large, the agreement with the results obtained from the
individual expansions is quite reasonable.
The deviations of the values $f(\Delta_1 ,\nu ,\eta)$
from the other resummation estimates might serve as a rough 
measure of the numerical accuracy achieved here.

The situation is less favorable for the second group of exponents,
$\eta_{\sf c}^*\equiv\eta_{\sf c}^{\text{sp}}(u^*),
\ldots,\Phi$, whose
estimates are given in Tables \ref{Tsp3} and \ref{Tsp4}.
Their series exhibit poor convergence properties.
One should be cautious in relying on estimates
derived from individual series expansions of an apparently
divergent nature, like in the case of the crossover
exponent $\Phi$.
In this group the exponent
$\alpha_1$ has the estimates
with the least scattering.
The best series convergence has $\alpha_1^{-1}$,
and the corresponding Pad\'e-Borel estimates are
$\alpha_1=0.342$ for $n=0$
and $\alpha_1=0.279$ for $n=1$.
Accepting these together with the
bulk values of $\nu$ and $\eta$ given above,
the estimates
of $\eta_{\sf c}^*$, $\alpha_{11}$, and $\Phi$ 
listed as $f(\alpha_1,\nu ,\eta)$
in the last columns of Tables \ref{Tsp3} and \ref{Tsp4}
were derived via scaling laws.

In order to check the quality of our procedure for estimating
the numerical values of critical exponents, we have applied the same
kind of analysis to the series expansions to second order 
of the bulk critical exponents. The results are shown in Table \ref{Tbce}
and compared with the estimates of Ref.\ \cite{lgz80} (last column).
The agreement gives us confidence in our method.

\begin{table}[hbtp]\squeezetable
\caption[truc]{\label{Tbce} Numerical estimates of bulk critical exponents. }
\begin{tabular}{c||d|d|d|d|d|d|d|d|d|d|d}
 & ${O_1/ O_2}$&${O_{1i}/ O_{2i}}$ & $[0/0]$ & $[1/0]$ & $[0/1]$ & $[2/0]$
& $[0/2]$ & $[1/1]$ & $R $ & $R_{inv}^{-1}$ & Ref.\ \cite{lgz80}\\
\hline\hline
\multicolumn{12}{c}{$n=0$}\\\hline
$\nu$ &$-$7.0&$-$4.1& 0.50& 0.602& 0.614
& 0.587& 0.583& 0.589& 0.590& 0.591& 0.588\\
$\gamma$ & $-$4.9&$-$2.5& 1.00& 1.204& 1.256
& 1.162& 1.137& 1.169& 1.172& 1.177& 1.1615\\
$\alpha$ & $-$7.0&6.2& 0.50& 0.194& 0.266
& 0.238& 0.238& 0.232& 0.230& --& 0.236\\
$\beta$ &$-$44.1&$-$13.6& 0.25& 0.301& 0.304
& 0.300& 0.300& 0.300& 0.300& 0.300& 0.302\\
$\Delta$ & $-$6.0&$-$2.7& 1.25& 1.505& 1.570
& 1.462& 1.434& 1.468& 1.470& 1.477& 1.462\\\hline
\multicolumn{12}{c}{$n=1$}\\\hline
$\nu$ &$-$27.7&$-$5.9& 0.50& 0.633& 0.654
& 0.628& 0.624& 0.628& 0.629& 0.630& 0.630\\
$\gamma$ &$-$11.3&$-$2.8& 1.00& 1.266& 1.363
& 1.243& 1.207& 1.244& 1.245& 1.254& 1.241\\
$\alpha$ &$-$27.7&2.8& 0.50& 0.101& 0.215
& 0.115& 0.148& 0.115& 0.114& --& 0.110\\
$\beta$ & 14.5&405.8& 0.25& 0.317& 0.321
& 0.321& 0.321& 0.321& --& --& 0.325\\
$\Delta$ &$-$17.5&$-$3.1& 1.25& 1.583& 1.703
& 1.564& 1.525& 1.565& 1.565& 1.575& 1.565\\\hline
\multicolumn{12}{c}{$n=2$}\\\hline
$\nu$ & 21.7&$-$9.1& 0.50& 0.656& 0.685
& 0.663& 0.661& 0.663& --& 0.664& 0.669\\
$\gamma$ & $\infty$&$-$3.2& 1.00& 1.312& 1.453
& 1.312& 1.273& 1.312&  --& 1.322& 1.316\\
$\alpha$ & 21.7&1.9& 0.50& 0.033& 0.181
& 0.011& 0.086& 0.010& --& --& $-$0.007\\
$\beta$ &  7.2&16.5& 0.25& 0.328& 0.334
& 0.339& 0.340& 0.340& --& --& 0.3455\\
$\Delta$ & 36.1&$-$3.5& 1.25& 1.640& 1.816
& 1.650& 1.609& 1.651& --& 1.662& 1.661\\\hline
\multicolumn{12}{c}{$n=3$}\\\hline
$\nu$ &  9.1&$-$15.9& 0.50& 0.673& 0.709
& 0.692& 0.693& 0.694& --& 0.695 & 0.705\\
$\gamma$ & 14.5&$-$3.6& 1.00& 1.346& 1.528
& 1.370& 1.333& 1.371& --& 1.382& 1.386\\
$\alpha$ &  9.1&1.6& 0.50& $-$0.019& 0.159
& $-$0.076& 0.042& $-$0.083& --& --& $-$0.115\\
$\beta$ &  5.2&9.5& 0.25& 0.336& 0.345
& 0.353& 0.356& 0.357& --& --& 0.3645\\
$\Delta$ &10.7&$-$4.0& 1.25& 1.682& 1.910
& 1.723& 1.686& 1.727& --& 1.739 & 1.751
\end{tabular}
\end{table}

The numerical values of
surface critical exponents gathered in
Tables \ref{Tsp1}-\ref{Tsp4}
generally are in reasonable agreement
both with previous estimates
based on the $\epsilon$
expansion as well as with those obtained by other means.
For comparisons we refer to Section III.8 of Ref.\ \cite{hwd},
where $\epsilon$ expansion estimates and
estimates that had been gained by alternative techniques
till 1985 are given,
and to Table \ref{Tsp5} for more recent results.
Note, however, that our estimates for the crossover exponent $\Phi$
are definitely lower than the values
$\Phi(n\!=\!1)\simeq 0.68$ and $\Phi(n\!=\!0)\simeq 0.67$
quoted in Ref.\ \cite{hwd}. The latter were
obtained by setting $\epsilon=1$ in the $\epsilon$ expansion of $\Phi$
to order $\epsilon^2$. On the other hand,
recent Monte Carlo simulations yielded the significantly lower
estimates $\Phi(1)=0.461\pm0.015$ \cite{rdww}, $\Phi(0)=0.530\pm0.007$
\cite{ml88}, and $\Phi(0)=0.496\pm0.005$ \cite{hg94}. Our present results
$\Phi(1)\simeq 0.54$ and $\Phi(0)\simeq 0.52$ are fairly close to
these values.

\begin{table}[hbtp]
\caption[truc]{\label{Tsp5}Monte Carlo estimates
for surface critical
exponents of the special transition in $d=3$ dimensions.}
\begin{tabular}{c||d|l}
\multicolumn{3}{c}{$n=0$}\\
\hline
$\gamma_{11}$&0.805(15) & Meirovitch \& Livne, 1988 \cite{ml88}\\
            &0.714(6) & Hegger \& Grassberger, 1994 \cite{hg94}\\
$\gamma_{1}$&1.304(6) & Meirovitch \& Livne, 1988 \cite{ml88}\\
           &1.230(2) & Hegger \& Grassberger, 1994 \cite{hg94}\\
$\Phi$       &0.530(7) & Meirovitch \& Livne, 1988 \cite{ml88}\\
            &0.496(4) & Hegger \& Grassberger, 1994 \cite{hg94}\\
\hline
\multicolumn{3}{c}{$n=1$}\\
\hline
$\beta_1      $&  0.18(2)&Landau \& Binder, 1990 \cite{lb90}\\
                &0.22& Vendruscolo {\it et al.}, 1992 \cite{vrf}\\
                &0.237(5)& Ruge {\it et al.}. 1993 \cite{rdww}\\
                &0.2375(15)& Ruge \&  Wagner 1995 \cite{rugew95}\\
$\gamma_{11}$&0.96(9) & Landau \& Binder, 1990 \cite{lb90}\\
                &0.788(1)& Ruge \& Wagner 1995 \cite{rugew95}\\
$\gamma_1$&1.41(14) &Landau \& Binder, 1990 \cite{lb90}\\ 
                &1.328(1)& Ruge \& Wagner 1995 \cite{rugew95}\\
$\Phi$& 0.59(4)&Landau \& Binder, 1990 \cite{lb90}\\
       & 0.74& Vendruscolo {\it et al.}, 1992 \cite{vrf}\\
       & 0.461(15)& Ruge {\it et al.}, 1993 \cite{rdww}
\end{tabular}
\end{table}

To see whether comparatively small estimates for $\Phi(n,d\!=\!3)$
can be obtained from the $\epsilon$ expansion, we have applied
the analogous summation techniques to the series
\begin{equation}\label{Phiexp}
2\Phi=1+a_1(n)\, \epsilon + a_2(n)\,\epsilon^2\;,
\end{equation}
whose coefficients are known to be  \cite{dd83,hwd}
\begin{equation}
a_1(n)=-{n+2\over 2(n+8)}=-\cases{\frac18 =0.125&for $n=0$\cr\cr
\frac16 \simeq 0.167&for $n=1$}
\end{equation}
and
\begin{eqnarray}
a_2(n)&=&{n+2\over 4(n+8)^3}\left[8\pi^2(n+8)-(n^2+35n+156)\right]\nonumber\\
&=&\cases{\frac{1}{16}\pi^2-\frac{39}{256}\simeq 0.4645&for $n=0$.\cr\cr
\frac{2}{27}\pi^2-\frac{16}{81}\simeq 0.5336 &for $n=1$.}
\end{eqnarray}
The results are shown in Table VI. It is reassuring
that the estimates obtained
via Pad\'e-Borel summation compare reasonably well
both with our above ones based on
the perturbation series at fixed $d=3$ as well
as with the Monte Carlo results mentioned. That
these estimates deviate considerably from
the values obtained from
the [2/0] approximant (\ref{Phiexp}) at  $\epsilon=1$ seems to be
due to the unusual largeness of the ${\cal O}(\epsilon^2)$ term
of $\Phi$. In summary, we conclude that the values of
the crossover exponent $\Phi(n,d)$
with $n=0,1$ and $d=3$ are indeed
significantly smaller than previously thought, being close to $0.5$.

\begin{table}[hbtp]
\caption[truc]{\label{Tsp6}Estimates for $\Phi(n,d\!=\!3)$ based on the $\epsilon$ expansion.}
\begin{tabular}{c||d|d|d|d|d|d|d|d|d|d}
 n& ${O_1/ O_2}$&${O_{1i}/ O_{2i}}$ & $[0/0]$ & $[1/0]$ & $[0/1]$ & $[2/0]$ & $[0/2]$ &
$[1/1]$ & $R $ & $R_i^{-1} $ \\\hline
$0$&        -0.27&-0.28& 0.5&0.438 & 0.444& 0.670
& 0.740& 0.487& 0.482& 0.483\\
1&-0.31&-0.33&0.5&0.417&0.429&0.683&0.757&0.480&0.474&0.475\\
\end{tabular}
\end{table}

An interesting aspect of  the above results
is worth mentioning: We may be quite confident that the inequality
\begin{equation}
\alpha^{\text{sp}}_{11}<0\label{allz}
\end{equation}
is satisfied for $d=3$ and $n=1$.
For one thing, our best numerical estimate
based on the massive RG approach at fixed $d=3$
is $\alpha^{\text{sp}}_{11}(n=1,3)\simeq -0.18$.
Second, the scaling relation (\ref{alpha11sl}) can be rewritten
at $d=3$ as
\begin{equation}
\alpha^{\text{sp}}_{11}(n,3)(n,d\!=\!3)=-2(\nu-\Phi)\;.
\end{equation}
In view of the various estimates given above it seems rather unlikely that
$\Phi(1,3)$ will be larger than $\nu(1,3)\simeq 0.630$, so that (\ref{allz}) should be
valid at $d=3$ and $n=1$,
a conclusion which may also be reached for $n=0$.

This has important consequences. As has been shown in
Ref.\ \cite{DN90a}, (\ref{allz}) plays
the role of an {\it irrelevance
criterion\/} for  {\it weak, short-range correlated randomness\/}
that couples to the {\it  surface energy density\/} (and
is restricted to the surface).
If it is satisfied, the fixed point describing
the special transition of the pure model is {\it stable} with respect to
this kind of  randomness., so that such
random ``surface-enhancement
disorder'' should be {\it irrelevant} at the special transition.
According to our numerical estimates, this irrelevance
should indeed apply. It has been verified by
Monte Carlo simulations recently \cite{PS98}.

\section{ORDINARY TRANSITION} \label{OrdTr}

In our analysis of the asymptotic critical behavior at the
special transition it turned out to be advantageous
to set the bare and renormalized surface enhancements
to their respective critical values $c_0=c_0^{\text{sp}}$
and $c=0$. The benefit was that we did not have to deal
with renormalization functions depending on {\it two\/} mass parameters
$m$ and $c$, a fact which facilitated the
computation of the required Feynman graphs considerably.

In the case of the ordinary transition we must study
the limit $c/m\to\infty$. For the sake of achieving
a similar simplification, it would be desirable to set $c=\infty$
(or $c_0=\infty$) from the outset. In doing so one is faced with
a known difficulty: Studying the functions $G^{(N,M)}$ with
$c_0=\infty$
does not easily give access to surface critical exponents via
the  RG equations of their renormalized analogs because
these bare functions as well as the renormalized ones with $c=\infty$
satisfy {\it Dirichlet boundary conditions\/}.
Fortunately it is known from
previous studies based on alternative RG
approaches \cite{DD80,DD81a,hwd,dde83}
how this problem can be overcome:
one must study the functions
\begin{equation}
{\cal G}^{(N,M)}(\bbox{x}_1,\ldots,\bbox{r}_M)\equiv
\left\langle\left[\prod_{j=1}^N\phi^{a_j}(\bbox{x}_j)\right]
\left[\prod_{k=1}^M \partial_n\phi^{b_k}(\bbox{r}_k)\right]
\right\rangle^{\!\!\text{cum}}\,,
\end{equation}
where $\partial_n$ means the derivative along the inner normal.
The functions ${\cal G}^{(N,M)}$ with $M>0$
do not vanish for $c_0=\infty$, and the scaling
dimension of $\partial_n\phi$ yields $\eta_{\|}^{\text{ord}}$,
the sole missing surface exponent.\footnote{Since the scaling dimension
$\Delta[\varepsilon_1]$
of the surface energy density $\varepsilon_1$
at the ordinary transition is exactly given
by $\Delta[\varepsilon_1]=d$, the analogs of (\ref{alpha1sl})
and (\ref{alpha11sl}) read
$\alpha_1^{\text{ord}}=\alpha -1$ and
$\alpha_{11}^{\text{ord}}=\alpha-2-\nu$, respectively \cite{dde83,DD81a}. The other
surface exponents are given by the scaling relations (\ref{allscal}).}

That the relevant information can be obtained from these
functions can be seen either by expanding the bare cumulants
$G^{(N,M)}$ in powers of $c_0^{-1}$ or
by noting that because of the Dirichlet boundary condition
$\partial_n\phi$ is the leading operator appearing in
the boundary operator expansion \cite{DD81a,hwd} of $\phi$.

\subsection{General considerations and the limit $c/m\to\infty$}  \label{genord}

Let us denote the  functions ${\cal G}^{(N,M)}$ with $c_0=\infty$
as ${\cal G}_\infty^{(N,M)}$.
Although we shall not present a complete analysis of the $c$-dependent
normalization conditions of Sec.~\ref{ssnc} and of the crossover from
special to ordinary surface critical behavior here, we will at least verify
that this renormalization procedure is consistent with the one
based on the ${\cal G}_\infty^{(N,M)}$, a scheme
whose results were briefly described in Ref.~\cite{Remprl} and which will
be exploited below.

We start by performing the mass renormalization
as described in Appendix \ref{amr}, and
introduce $\hat\sigma(p;m,c_0)$,
the analog of $\hat\sigma_0(p)$, via
\begin{equation}\label{sigmap}
\hat{G}^{(0,2)}[p;m_0(m),u_0,c_0]=
\left[\kappa+c_0-\hat{\sigma}(p;m,c_0)
\right]^{-1}\;.
\end{equation}
Its expansion to order $u_0^2$ is given in (\ref{sigmapr}) of
Appendix \ref{amr}.

Assuming that
the renormalized surface enhancement $c$
has an arbitrary value $0\le c<\infty$,
we imagine that the surface-enhancement renormalization
has been carried out in the way explained in Appendix \ref{asr}.
Substituting the resulting form (\ref{xxt}) of
$\big[\hat G^{(0,2)}(p)\big]^{-1}$ into
(\ref{zsp}) yields
\begin{equation}\label{62}
\big[Z_1(u,c/m)Z_\phi(u)\big]^{-1}-1=-
\lim_{p \rightarrow0}{m \over p}{\partial \over \partial p}\,
\big[\hat\sigma(p;m,c+\delta c)
-\hat\sigma(0;m,c+\delta c)
\big]\,.
\end{equation}

We wish to study what happens to the perturbation expansion in $u$
of the right-hand side of (\ref{62}) in the limit $c/m\to\infty$.
To this end, we set $m=1$ and let $c\to\infty$.
Then the free propagator --- namely
(\ref{freeprop}), with $c_0$ and $\kappa_0$
replaced by $c$ and $\kappa$, respectively ---
goes over into the Dirichlet propagator (\ref{dirprop}).
Further, the perturbative corrections caused by
the shift $\delta c$ to the term inside
the square brackets of (\ref{62})
vanish as $c\to\infty$.%
\footnote{A simple way to see this is to note that such
corrections involve free propagators with points on the surface.
Dimensional arguments lead to the same conclusion.} %
Hence we have
\begin{equation}\label{sigmaD}
\lim_{c\to\infty}\big[\hat\sigma(p;m,c+\delta c)
-\hat\sigma(0;m,c+\delta c)
\big]=\hat\sigma^{(\text{D})}(p;m)
-\hat\sigma^{(\text{D})}(0;m)
\end{equation}
with $\hat\sigma^{(\text{D})}(p;m)=\hat\sigma(p;m,\infty)$.
As we have seen, the graphs of $\hat\sigma^{(\text{D})}(p;m)$
are obtained from  those of $\hat\sigma(p;m,c)$ by associating with
all full lines the Dirichlet propagator $\hat G_{\text{D}}$ rather
than the $c$-dependent one (\ref{freeprop}), and with all full
lines labeled ``s'' (cf.\ Appendix \ref{amr}) its
surface part, given by (\ref{gs}) with $c_0=\infty$.
Note that  these graphs
are not in general uv finite at $d=3$.
But subtraction of their values at $p=0$,
which is provided by the last term in (\ref{sigmaD}),
is sufficient to make them so. In other words,
in the limit $c\to \infty$, surface-enhancement
renormalization reduces to
an additive renormalization.

To see how this relates to our approach based on
the $c_0=\infty$ functions ${\cal G}_\infty^{(N,M)}$,
we return to the representation (\ref{sigmaexpans}) of $\hat\sigma_0$
in terms of the self-energy $\hat\Sigma$.
Since the denominator of the fraction in (\ref{sigmaexpans})
becomes one for $c_0=\infty$, we have
$\hat\sigma^{(\text{D})}=\hat{g}^{\text{T}}\hat{T}[ G_{\text{D}}]\hat{g}$,
where $T[G]$ is the T-matrix introduced in (\ref{Tmatrix}).
Now the reduced
propagator (\ref{redprop}) can be written as
\begin{equation}
\hat g(p;z')=e^{-\kappa z'}={\partial\over\partial z}\,
\hat G_{\text{D}}(\bbox{p};z,z')\bigg|_{z=0}\;.
\end{equation}
Thus we get
\begin{equation}\label{Ginfty}
\hat{\cal G}_\infty^{(0,2)}[p;m_0(m)]=-\kappa +
\hat\sigma^{(\text{D})}(p;m)\;,
\end{equation}
where it should be remembered \cite{Syman,hwd,dde83} that
$\partial_z\partial_{z'}\hat G_{\text{D}}(p;z,z')$
has a contribution of the form $[-\delta (z-z')]$;
we have dropped the implied singularity $[-\delta(0)]$
in the zero-loop term $(-\kappa)$, interpreting
$\partial_n\hat G_{\text{D}} \loarrow{\partial_n}$ as
the limit of $\partial_z\partial_{z'}\hat G_{\text{D}}(p;z,z')$
as $z,z'\to 0$ with $z\ne z'$.
Combining these findings with (\ref{62}) and (\ref{sigmaD}) yields
\begin{equation}\label{zino}
[Z_1(u,\infty)Z_\phi(u)]^{-1}=-
\lim_{p \to 0}{m \over p}{\partial \over \partial p}
\left[\hat{\cal G}_\infty^{(0,2)}(p)
-\hat{\cal G}_\infty^{(0,2)}(0)\right]\;.
\end{equation}

Next, let us recapitulate our renormalization scheme
for the ${\cal G}_\infty^{(N,M)}$ \cite{Remprl}.
Aside from the previous bulk renormalization functions, it
involves a renormalization factor $Z_{1,\infty}(u)$, which enters
the definition of the renormalized surface operator:
\begin{equation}
(\partial_n\phi)_{\text{ren}}=[Z_{1,\infty}
Z_\phi]^{-1/2}\,\partial_n\phi \;,
\end{equation}
and of the renormalized functions:
\begin{equation}\label{Ginftyr}
\hat {\cal G}_{\infty,\text{ren}}^{(N,M)}(\{\bbox{p}\};\{z_j\};m,u)
=Z_\phi^{-(N+M)/2}Z_{1,\infty}^{-M/2}
\left[\hat {\cal G}_\infty^{(N,M)}(\{\bbox{p}\};\{z_j\})
-\delta_{N,0}^{M,2}\,\hat {\cal G}_\infty^{(0,2)}(0)\right].
\end{equation}
One evident normalization condition is
\begin{mathletters}\begin{equation}
\hat {\cal G}_{\infty,\text{ren}}^{(0,2)}(0;m,u)=0\;.
\end{equation}
The other, 
\begin{equation}
{\partial\over\partial p^2}\,\left.
\hat {\cal G}_{\infty,\text{ren}}^{(0,2)}(p;m,u)\right|_{p=0}=
-{1\over 2m}\;,
\end{equation}\end{mathletters}\noindent
(suggested by the corresponding zero-loop result)
serves to fix $Z_{1,\infty}$. In conjunction with (\ref{Ginftyr})
it implies the relation:
\begin{equation}\label{z11}
Z_{1,\infty}(u)Z_\phi(u)=
-\lim_{p\to 0}{m\over p}{\partial\over\partial p}
\left[\hat{\cal G}_\infty^{(0,2)}(p)
-\hat{\cal G}_\infty^{(0,2)}(0)\right],
\end{equation}
whose comparison with (\ref{zino}) reveals that
\begin{equation}\label{Z1Zphilarge}
\Big[Z_1(u,\infty)Z_\phi(u)\Big]^{-1}=\lim_{c/m\to\infty}
\Big[Z_1(u,c/m)Z_\phi(u)\Big]^{-1}=Z_{1,\infty}(u)Z_\phi(u)
\end{equation}
to any order of perturbation theory.

We introduce the analog of the exponent function $\eta_1^{\text{sp}}$
by
\begin{equation}\label{eta1infty}
\eta_{1,\infty}(u)\equiv\tilde\eta_{1,\infty}(\tilde u)\equiv
\left. m{\partial\over\partial m}
\right|_0\,\ln Z_{1,\infty}(u)=
\beta(u)\,{\partial
\ln Z_{1,\infty}(u)\over\partial u} \; ,
\end{equation}
where $\tilde u$ is the rescaled coupling constant of (\ref{utildedef}).
The fixed-point value of this function, $\eta_{1,\infty}(u^*)$, is related
to $\eta_{\|}^{\text{ord}}$
via \cite{Remprl} (cf.\ Ref.\ \cite{hwd}) 
\begin{equation}\label{etaparord}
\eta_\parallel^{\text{ord}}=2+\eta_1^{\text{ord}}(u^*)+\eta_\phi(u^*)\;,
\end{equation}
as we shall verify below. Reasoning in a standard fashion, we find that
\begin{equation}\label{Z1inftyasf}
Z_{1,\infty}\sim(u-u^*)^{\eta_{1,\infty}(u^*)/\omega}
\sim m^{\eta_{1,\infty}(u^*)}
\end{equation}
as $m\to 0$ (or $u\to u^*$), with fixed
bare interaction constant $u_0$
(and $c_0=\infty$).

The renormalized functions
$G_{\infty,\text{ren}}^{(N,M)}$
satisfy the analog of the CS equation (\ref{CSE}):
\begin{equation}\label{CSEinfty}
\left[m{\partial\over\partial m}
+\beta(u){\partial\over\partial u}
+{{N+M}\over 2}\,\eta_\phi(u)+{M\over 2}\,\eta_{1,\infty}(u)\right]
G_{\infty,\text{ren}}^{(N,M)}(;m,u)=\Delta G_{\infty,\text{ren}}\,,
\end{equation}
in which the inhomogeneous term
$\Delta G_{\infty,\text{ren}}$ is defined just as $\Delta G_{\text{ren}}$ in
(\ref{inhomo}), but with
$G_{\text{ren,sp}}^{(N,M;1,0)}$ replaced by
$G_{\infty,\text{ren}}^{(N,M;1,0)}$, the corresponding cumulant with an
insertion of $\frac12 \int_V \phi^2$. Neglecting $\Delta G_{\infty,\text{ren}}$,
we can exploit in the usual fashion
the resulting homogeneous CS equation together with
the asymptotic forms (\ref{Zphiasform}) and (\ref{Z1inftyasf})
of $Z_\phi$ and $Z_{1,\infty}$ to conclude that the bare cumulants
behave as
\begin{equation}\label{scformord}
G_\infty^{(N,M)}(\bbox{x},\bbox{r};m_0,u_0)\sim
m^{(N\beta+M\beta_1^{\text{ord}})/\nu}\,
\Psi_\infty^{(N,M)}\!\left(m\bbox{x},m\bbox{r}\right)
\end{equation}
near criticality. That these scaling forms carry over to
the asymptotic behavior of the
functions $G^{(N,M)}(\bbox{x},\bbox{r};m_0,u_0,c_0)$
near the ordinary transition can be seen in the ways mentioned
in the introduction to this section and expounded in Refs.\ \cite{hwd}
(use either the expansion of the bare functions in powers of $1/c_0$
or the boundary operator expansion).
Here we shall present an alternative derivation, which
is based directly on our $c$-dependent renormalization scheme.

First, we need the asymptotic scale dependence of the
variable $c=c(m)$ near the ordinary fixed point. This can be
conveniently obtained from the reformulated normalization
condition (\ref{cr}). The bare function $G^{(0,2)}(p\!=\!0)=\chi_{11}$
approaches a finite value $\chi_{11}^{\text{ord}}(u_0,c_0,\Lambda)$
as $T\to T_c$ ($m\to 0$) with fixed $u_0$ and $c_0<c_0^{\text{sp}}$.
Using the  limiting behavior (\ref{Z1Zphilarge}) of $Z_\phi Z_1$
for $c/m\to\infty$ together
with the asymptotic forms (\ref{Zphiasform}) and (\ref{Z1inftyasf}) of 
$Z_\phi$ and $Z_{1,\infty}$, we arrive at the relation
\begin{equation}
(c+m)\,m^{\eta+\eta_{1,\infty}(u^*)}\sim \chi_{11}^{\text{ord}}\;,
\end{equation}
which yields
\begin{equation}
c\sim m^{-(\eta_\|^{\text{ord}}-2)}\:.
\end{equation}

The second ingredient we shall need is the asymptotic
behavior of the dimensionless function
\begin{equation}
m^{-(N+M)(d-2)/2}\,
G^{(N,M)}_{\text{ren}}(\bbox{x},\bbox{r};m,u,c)=
G^{(N,M)}_{\text{ren}}(m\bbox{x},m\bbox{r};1,u,c/m)
\end{equation}
as ${\sf c}\equiv c/m\to\infty$.
Based on our knowledge of the $1/c_0$ expansion
(cf.\ the analogous
considerations in Sec.\ III C 6 of Ref.\ \cite{hwd}),
we anticipate that
\begin{equation}
G^{(N,M)}_{\text{ren}}(\bbox{x},\bbox{r};1,u,{\sf c})\;
\mathop{\approx}\limits_{{\sf c}\to \infty}\;
 {\sf c}^{-M}\,{\cal F}^{(N,M)}_\infty (\bbox{x},\bbox{r},u)
+{\sf c}^{-1}\,{\cal R}(u)\,\delta_{N,0}^{M,2}\,
\delta (\bbox{r}_{12})\;,
\end{equation}
where $\bbox{r}_{12}=\bbox{r}_1-\bbox{r}_2$.
When these results are inserted into
$G^{(N,M)} = Z_\phi^{(N+M)/ 2}Z_1^{M/2}
 G_{\text{ren}}^{(N,M)}$, each one of the $M$ surface operators $\phi_s$
is found to contribute a factor
\begin{equation}
m^{(d-2)/2}\left[Z_\phi (u)\,Z_{1,\infty}(u)\right]^{-1/2}
 (c/m)^{-1}
\sim m^{(d-2+\eta_\|^{\text{ord}})/2}
\end{equation}
to the prefactor of ${\cal F}^{(N,M)}_\infty$. Hence we recover indeed
the familiar scaling form [cf.\ (\ref{scformord})]:
\begin{equation}\label{scford2}
G^{(N,M)}(\bbox{x},\bbox{r})\sim
m^{(N\beta+M\beta_1^{\text{ord}})/\nu}\,
{\cal F}^{(N,M)}_\infty (m\bbox{x},m\bbox{r},u^*)\;.
\end{equation}
In the special case $(N,M)=(0,2)$, a contribution
$m^{(\eta_\|^{\text{ord}}-1)}\,
{\cal R}(u^*)\,\delta (\bbox{r}_{12})$ and similar subleading
ones $\propto \delta (\bbox{r}_{12})$ appear,
which we have suppressed in (\ref{scford2}).

\subsection{Results of perturbation theory to two-loop order}\label{sefo}

We now turn to the explicit calculation
of the renormalization factor
$Z_{1,\infty}(u)Z_\phi(u)$ up to two-loop order,
restricting ourselves again to the case $d=3$.

Setting $c_0=\infty$ in (\ref{sigmapr})
gives us the perturbation expansion of $\hat\sigma^{(\text{D})}$
to second order in $u_0$. This we insert into (\ref{Ginfty}),
and the so-obtained form of
$\hat{\cal G}_{\infty}^{(0,2)}$ then into (\ref{z11}).
There are two simplifying features we can benefit from. First,
the one-loop graph of $\hat\sigma^{(\text{D})}$ differs from its $c=0$
counterpart by a minus sign. This means that the term linear in $u_0$
agrees with its counterpart for $Z_1^{\text{sp}}Z_\phi$.
Second, as we shall show in Appendix \ref{appord},
the contributions from the two-loop graphs (3) and (4)
of Fig.\ 3 cancel. Hence we get
\def\epsfsize#1#2{0.6#1}
\begin{eqnarray}\label{z2l}
Z_{1,\infty}Z_\phi&=&1+{n+2\over12}\, \frac{u_0}{8\pi m}
-2m\left.\frac{\partial}{\partial p^2}\right|_{p=0}\Biggl\{\raisebox{-1.0em}{\epsfbox{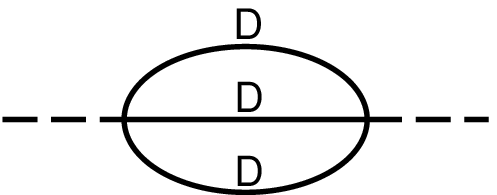}}\;
\nonumber\\[1em]
&&\mbox{}
-\frac{1}{2\kappa}{n+2 \over 18}\,u_0^2 \left[ I_2(m^2)-
m^2I_3(m^2)\right]\Biggr\}+{\cal O}\big(u_0^3\big)\;.
\end{eqnarray}
The required $u_0^2$ term is easily calculated (see
Appendix \ref{appord}). One finds
\begin{mathletters}
\begin{equation}\label{zc0} 
Z_{1,\infty}Z_\phi=1+{n+2\over12}\,\frac{u_0}{8\pi m}
+{n+2\over3}\left(\frac{u_0}{8\pi m}\right)^2 \,C
+{\cal O}\big(u_0^3\big)
\end{equation}
with
\begin{equation}\label{co}
C={107\over 162}-  {7\over 3} \ln{4\over 3}-0.094299 =-0.105063 \;.
\end{equation}\label{allzc0}
\end{mathletters}\noindent
Upon expressing $u_0$ in terms
of the rescaled renormalized coupling constant
$\tilde u=u\,(n+8) /48m\pi$
[cf.\ (\ref{utildedef})], the result becomes
\begin{equation}
Z_{1,\infty}(u)Z_\phi(u)=1+{n+2\over 2(n+8)}\,\tilde{u}
+{12\,(n+2)\over (n+8)^2}
\left(C+{n+8\over 24}\right)\tilde{u}^2 +{\cal O}\big(u^3\big)\;.
\label{zvr1} \end{equation}
From it the exponent function appearing on the right-hand side of (\ref{etaparord})
can be deduced in a straightforward fashion. One obtains
\begin{equation}\label{eg} 
\eta_\parallel^{\text{ord}}(u)=
\tilde{\eta}_\parallel^{\text{ord}}(\tilde{u})=
2-{n+2\over 2(n+8)}\,\tilde{u}
- {24\,(n+2)\over (n+8)^2}\left(C+ {n+14\over 96}\right)\tilde{u}^2
+{\cal O}\big(u^3\big)\,.
\end{equation}
The corresponding series expansions of the surface exponents
$\Delta^{\text{ord}}_1$, $\eta^{\text{ord}}_\perp$, $\beta^{\text{ord}}_1$,
$\gamma^{\text{ord}}_{11}$, $\gamma^{\text{ord}}_1$,
$\delta^{\text{ord}}_1$, and $\delta^{\text{ord}}_{11}$
follow again by substituting  (\ref{eg})
together with the expansions (\ref{nuetaexp})  of  $\nu$ and $\eta$ into
the scaling-law expressions (\ref{allscal}).

\subsection{Numerical estimates for the %
surface critical exponents of the ordinary transition}

Following the strategy described in Sec.\ \ref{SCESpTr}, one
can analyze the above power series for the critical
exponents of the ordinary transition and extract
numerical estimates. The results are
shown in Tables \ref{To0}-\ref{To3}, where the
entries have the same meaning
as in Tables \ref{Tsp1}--\ref{Tsp2} (Sec.\ \ref{SCESpTr}).
As before, the fixed-point values
$u^*(n)$ of Refs.\ \cite{bgmn} and \cite{jug}, obtained by
Pad\'e-Borel resummation of the two-loop result for
the $\beta$ function, were used.

\begin{table}[hbtp]
\caption[truc]{\label{To0}Surface critical exponents %
of the ordinary transition for $d=3$ and $n=0$. As %
fixed-point value we used $\tilde{u}^*=1.632$.}
\begin{tabular}{c||d|d|d|d|d|d|d|d|d}
 & ${O_1/O_2}$&${O_{1i}/O_{2i}}$ & $[0/0]$ & $[1/0]$ & $[0/1]$ & $[2/0]$ & $[0/2]$
&$[1/1]$ & $f(\eta_{\|},\nu,\eta)$\\
\hline\hline
$\eta_{\|}$ &  2.5&2.0& 2.00& 1.796& 1.815
& 1.715& 1.734& 1.660& 1.660\\
$\Delta_1$ &  4.4&7.8& 0.25& 0.352& 0.364
& 0.375& 0.380& 0.382& 0.394\\
$\eta_\perp$ &  3.6&2.6& 1.00& 0.898& 0.907
& 0.870& 0.877& 0.859& 0.843\\
$\beta_1$ & $-$1.9&$-$1.6& 0.75& 0.852& 0.864
& 0.799& 0.790& 0.817& 0.782\\
$\gamma_{11}$ &  0.0&0.0& $-$0.50& $-$0.500& $-$0.500
& $-$0.424& $-$0.434&  -& $-$0.388\\
$\gamma_1$ & 15.4&$-$11.4& 0.50& 0.653& 0.681
& 0.663& 0.662& 0.664& 0.680\\
$\delta_1$ &  2.5&3.1& 1.67& 1.780& 1.788
& 1.825& 1.832& 1.854& 1.870\\
$\delta_{11}$ &  2.1&2.7& 0.33& 0.424& 0.433
& 0.466& 0.476& 0.504& 0.504
\end{tabular}
\end{table}

\begin{table}[hbtp]
\caption[truc]{\label{To1}Surface critical exponents %
of the ordinary transition for $d=3$ and $n=1$. As %
fixed-point value we used $\tilde{u}^*=1.597$.}
\begin{tabular}{c||d|d|d|d|d|d|d|d|d}
 & ${O_1/O_2}$&${O_{1i}/O_{2i}}$ & $[0/0]$ & $[1/0]$ & $[0/1]$ & $[2/0]$ & $[0/2]$
&$[1/1]$ & $f(\eta_{\parallel},\nu,\eta)$\\
\hline\hline
$\eta_{\|}$ &  2.3&1.8& 2.00& 1.734& 1.765
& 1.618& 1.655& 1.528& 1.528\\
$\Delta_1$ & 3.0&5.0& 0.25& 0.383& 0.404
& 0.427& 0.440& 0.450& 0.464\\
$\eta_\perp$ &  3.0&2.2& 1.00& 0.867& 0.883
& 0.823& 0.837& 0.801& 0.779\\
$\beta_1$ & $-$2.5&$-$1.9& 0.75& 0.883& 0.904
& 0.829& 0.815& 0.845& 0.796\\
$\gamma_{11}$ &  0.0&0.0& $-$0.50& $-$0.500& $-$0.500
& $-$0.402& $-$0.418&  -& $-$0.333\\
$\gamma_1$ &  5.7&$-$40.4& 0.50& 0.700& 0.749
& 0.735& 0.742& 0.742& 0.769\\
$\delta_1$ &  2.2&2.7& 1.67& 1.815& 1.829
& 1.883& 1.898& 1.941& 1.966\\
$\delta_{11}$ &  1.9&2.5& 0.33& 0.452& 0.468
& 0.514& 0.533& 0.582& 0.582
\end{tabular}
\end{table}

\begin{table}[hbtp]
\caption[truc]{\label{To2}Surface critical exponents %
of the ordinary transition for $d=3$ and $n=2$. As %
fixed-point value we used $\tilde{u}^*=1.558$.}
\begin{tabular}{c||d|d|d|d|d|d|d|d|d}
 & ${O_1/O_2}$&${O_{1i}/O_{2i}}$ & $[0/0]$ & $[1/0]$ & $[0/1]$ & $[2/0]$ & $[0/2]$
&$[1/1]$ & $f(\eta_{\parallel},\nu,\eta)$\\
\hline\hline
$\eta_{\|}$ & 2.2&1.6& 2.00& 1.688& 1.730
& 1.545& 1.598& 1.422& 1.422\\
$\Delta_1$ &  2.4&3.9& 0.25& 0.406& 0.435
& 0.470& 0.493& 0.514& 0.528\\
$\eta_\perp$ &  2.7&1.9& 1.00& 0.844& 0.865
& 0.787& 0.808& 0.753& 0.727\\
$\beta_1$ & $-$3.2&$-$2.1& 0.75& 0.906& 0.935
& 0.856& 0.840& 0.868& 0.810\\
$\gamma_{11}$ &  0.0&0.0& $-$0.50& $-$0.500& $-$0.500
& $-$0.387& $-$0.408& -& $-$0.282\\
$\gamma_1$ &  3.9&42.1& 0.50& 0.734& 0.805
& 0.794& 0.814& 0.815& 0.851\\
$\delta_1$ &  2.0&2.5& 1.67& 1.840& 1.860
& 1.928& 1.952& 2.019& 2.051\\
$\delta_{11}$ &  1.8&2.3& 0.33& 0.472& 0.494
& 0.550& 0.579& 0.651& 0.652
\end{tabular}
\end{table}

\begin{table}[hbtp]
\caption[truc]{\label{To3}Surface critical exponents %
of the ordinary transition for $d=3$ and $n=3$. As %
fixed-point value we used $\tilde{u}^* = 1.521$.}
\begin{tabular}{c||d|d|d|d|d|d|d|d|d}
 & ${O_1/O_2}$&${O_{1i}/O_{2i}}$ & $[0/0]$ & $[1/0]$ & $[0/1]$ & $[2/0]$ & $[0/2]$
&$[1/1]$ & $f(\eta_{\parallel},\nu,\eta)$\\
\hline\hline
$\eta_{\|}$ &  2.1&1.5& 2.00& 1.654& 1.705
& 1.489& 1.556& 1.338& 1.338\\
$\Delta_1$ &  2.1&3.4& 0.25& 0.423& 0.459
& 0.504& 0.538& 0.574& 0.586\\
$\eta_\perp$ &  2.5&1.8& 1.00& 0.827& 0.853
& 0.759& 0.787& 0.714& 0.685\\
$\beta_1$ & $-$4.1&$-$2.4& 0.75& 0.923& 0.959
& 0.880& 0.862& 0.889& 0.824\\
$\gamma_{11}$ &  0.0&0.0& $-$0.50& $-$0.500& $-$0.500
& $-$0.377& $-$0.401&  -& $-$0.238\\
$\gamma_1$ &  3.1&16.3& 0.50& 0.759& 0.850
& 0.842& 0.880& 0.882& 0.927\\
$\delta_1$ &  1.8&2.3& 1.67& 1.859& 1.884
& 1.963& 1.995& 2.088& 2.124\\
$\delta_{11}$ &  1.7&2.3& 0.33& 0.487& 0.515
& 0.578& 0.617& 0.711& 0.711
\end{tabular}
\end{table}

For most of the obtained truncated series expansions, the coefficients
do not alter in sign, and the truncated series of their reciprocal (i.e., their
`inverse series')
display a similar behavior.
Of this kind are the series for 
$\eta^{\text{ord}}_{\parallel}$, $\Delta^{\text{ord}}_1$,
$\eta^{\text{ord}}_\perp$, $\delta^{\text{ord}}_1$,
and $\delta^{\text{ord}}_{11}$ with $n=0,\ldots,3$,
and for $\gamma^{\text{ord}}_1$ with $n=2$ and $3$.
Let $s_{[p/q]}$, with $p+q\le 2$, be the values resulting from
Pad\'e approximants of type $[p/q]$ (and listed in the columns
marked $[p/q]$), and let $s_p\equiv s_{[p/0]}$.
Looking at Tables \ref{To0}--\ref{To3} one realizes that
the sequences of values $s_{[p/q]}$ associated with each
one of these critical indices have the following feature:
The values move away from $s_0$ such that the second-order
approximants $[p/(2-p)]$
give values farther away from
$s_0$ than the first-order ones $[p/(1-p)]$ and that furthermore
$s_{[1/1]}$ is the most distant one. In other words, either they
increase according to
\begin{equation}\label{ssimonot}
s_0 <s_1 <s_2 <s_{[1/1]}
\quad\mbox{and}\quad
s_0 <s_{[0/1]} <s_{[0/2]} <s_{[1/1]}
\end{equation}
or else they decrease in the corresponding fashion. In most cases
even the stronger chain of inequalities
\begin{equation}\label{strongmonot}
s_0 <\min\big\{s_1,s_{[0/1]}\big\} <
\max\big\{s_1,s_{[0/1]}\big\}<
\min\big\{s_2,s_{[0/2]}\big\}<
\max\big\{s_2,s_{[0/2]}\big\}  <s_{[1/1]}
\end{equation}
or its decreasing analog applies.

The value $s_{[1/1]}$ always comes last in these sequences.
Using it to extrapolate the series amounts to anticipating that
the next (thus far unknown) terms of the power series
expansion in $\tilde u$
have coefficients of the same sign. This assumption might well be true
for some of the series (an example of this kind is the bulk
exponent $\eta$), and in view of the just mentioned feature of the
$s_{[p/q]}$ with $p+q\le 2$ it seems legitimate to us to accept it.
Accordingly we consider $s_{[1/1]}$ to be the
best among all those estimates $s_{[p/q]}$ with $p+q\le 2$
for a given exponent that we obtained from its individual
series expansion.

For $\gamma^{\text{ord}}_1$
with $n=0$ and $1$, only the first chain of inequalities of
(\ref{ssimonot}) holds. Its inverse series has
first-order and second-order
corrections of different signs, and hence may be treated
by the Pad\'e-Borel method. The resulting resummation values
(the analog of
the ones denoted $R_i^{-1}$ in Tables \ref{Tsp1}--\ref{Tsp4})
agree with $s_{[1/1]}$ up to three decimals.

In the case of $\beta^{\text{ord}}_1$ (with $n=0,1,2,3$)
both the direct and the inverse series
are alternating. The results of our resummations differ from
the values of the $[1/1]$ approximants only in the third decimal.
(Therefore we have not listed them separately.)	
The series for $\gamma^{\text{ord}}_{11}$ have
zero first-order corrections and hence are not
well adapted for estimating this critical exponent.

In order to gain further improved
estimates, we follow a similar strategy as we did
in Sec.\ \ref{SCESpTr} when estimating the critical exponents
of the special transition: we try to exploit the above results
in conjunction with the available high-precision estimates
for the bulk exponents $\nu$ and $\eta$. To this end we
substitute our $[1/1]$ values for
$\eta^{\text{ord}}_{\parallel}$,
together with the estimates
taken from Ref.\ \cite{lgz80},
$\nu=0.588$, $\eta=0.027$
(for $n=0$),
$\nu=0.630$, $\eta=0.031$ (for $n=1$),
$\nu=0.669$, $\eta=0.033$ (for $n=2$), 
and $\nu=0.705$, $\eta=0.033$ (for $n=3$),
into the scaling-law expressions (\ref{allscal}).
The results are given as $f(\eta_{\parallel},\nu,\eta)$
in the last row of Tables \ref{To0}-\ref{To3}.
As one sees, in those cases in which
the Pad\'e values $s_{[p/q]}$ move away
from $s_0$ in a given direction such that either
(\ref{ssimonot})  --- or even (\ref{strongmonot}) --- or else
their corresponding decreasing analogs hold,
the estimates $f(\eta_{\parallel},\nu,\eta)$ turn out to be displaced even
further in the same direction.

We consider our estimates $f(\eta_{\parallel} ,\nu ,\eta)$
as the best we could attain from the available
knowledge on the series expansions,
within the present approximation scheme.
In some cases they differ significantly
from the zeroth-order values $s_0$ we started
from. Like in the case of the special transition,
our best estimates agree
reasonably well both with the earlier ones based on
the $\epsilon$ expansion \cite{hwd,DN86}
as well as with more recent
computer-simulation results
\cite{ml88,hg94,PS98,dblw90,dbl85,ko,lpb89}.
The latter are gathered in
Table \ref{To5}. For references to earlier
numerical estimates and their comparison with
$\epsilon$-expansion results, the reader may
consult Ref.\ \cite{hwd}.

\begin{table}[hbtp]
\caption[truc]{\label{To5}Monte Carlo estimates for  the surface critical
exponents of the ordinary transition in $d=3$ dimensions.}
\begin{tabular}{c|l|l}
\multicolumn{3}{c}{$n=0$}\\
\hline
$\gamma_{11}$&$-0.38(2)$& Meirovitch \& Livne, 1988 \cite{ml88}\\
            &$-0.353(17)$ &De'Bell {\it et al.}, 1990 \cite{dblw90}\\
            &$-0.383(5) $& Hegger \& Grassberger, 1994 \cite{hg94}\\
$\gamma_{1}$&$+0.718(8)$ & De'Bell \& Lookman, 1985 \cite{dbl85}\\
            &$+0.687(5)$ & Meirovitch \& Livne, 1988 \cite{ml88}\\
            &$+0.694(4)$ & De'Bell {\it et al.}, 1990 \cite{dblw90}\\
           &$+0.679(2) $& Hegger \& Grassberger, 1994 \cite{hg94}\\
\hline
\multicolumn{3}{c}{$n=1$}\\
\hline
$\beta_1      $&$+  0.79(2)$&Kikuchi \& Okabe, 1985 \cite{ko}\\
               & $+0.78(2)$&Landau \& Binder, 1990 \cite{lb90}\\
               &$+0.75(2)$& Ruge {\it et al.}, 1993 \cite{rdww}\\
               &$+0.807(4)$& Ruge \& Wagner, 1995 \cite{rugew95}\\
               &$+0.80 \pm 0.01$&Pleimling \& Selke, 1998 \cite{PS98}\\
$\gamma_1$&$+0.78(6)$ &Landau \& Binder, 1990 \cite{lb90}\\ 
                &$+0.760(4)$& Ruge \& Wagner, 1995 \cite{rugew95}\\
&$+0.78 \pm 0.05$&Pleimling \& Selke, 1998 \cite{PS98}\\

$\gamma_{11}$&$-0.25\pm 0.1$&Pleimling \& Selke, 1998 \cite{PS98}\\
$\delta_1      $&$+ 2.00(8)$&Kikuchi \& Okabe, 1985 \cite{ko}\\
\hline
\multicolumn{3}{c}{$n=2$}\\
\hline
$\beta_1      $&$+  0.84$&Landau {\it et al.}, 1989 \cite{lpb89}\\
$\gamma_1$& $\simeq  {2\over3}$ &Landau {\it et al.}, 1989 \cite{lpb89}\\
\end{tabular}
\end{table}

Specifically, our estimates
$\gamma^{\text{ord}}_{11}(n\!=\!0)\simeq -0.388$
and $\gamma^{\text{ord}}_1(n\!=\!0)\simeq 0.680$ for
the polymer universality class ($n=0$)
are in excellent agreement with
the recent (apparently very precise) Monte Carlo estimates
$\gamma^{\text{ord}}_{11}(n\!=\!0)=-0.383(5)$
and $\gamma^{\text{ord}}_1(n\!=\!0)=0.679(2)$
by Hegger and Grassberger \cite{hg94}.
Likewise for the Ising universality class, our
numerical values $\beta^{\text{ord}}_1(n\!=\!1)\simeq 0.80$
and $\gamma^{\text{ord}}_1(n\!=\!1)\simeq 0.77$
are very close to the Monte Carlo estimates
$\beta^{\text{ord}}_1(n\!=\!1)=0.807(4)$
and $\gamma^{\text{ord}}_1(n\!=\!1) = 0.760(4)$
of Ruge {\it et al.}\, \cite{rugew95}. 
Landau and Binder's earlier Monte Carlo estimates \cite{lb90}
$\beta^{\text{ord}}_1(n\!=\!1)\simeq 0.78$
and $\gamma^{\text{ord}}_1(n\!=\!1) = 0.78(6)$
are slightly smaller and larger, respectively.
The more recent ones by Pleimling and Selke \cite{PS98}
coincide within their error bars with those of Ref.\ \cite{rugew95}
and our best estimate.

There also exist some experimental results%
\footnote{For background and a review of experimental work on surface
critical  behavior, see
Ref.\ \cite{Dos92}.}
 with which these
theoretical Ising values can be compared.
Sigl and Fenzl \cite{sfexp} were able to extract the value
$\beta_1 =0.83\pm 0.05$ from capillary-rise experiments
on the transition from partial to complete
wetting in critical mixtures of lutidine and water with different
amounts of dissolved potassium chloride. Using the technique of
x-ray scattering at grazing incidence
\cite{DW83,DW84,Dos92,DH95},
Mail{\"a}nder {\it et al.}\ \cite{dmjp} investigated
the surface critical behavior of a FeAl alloy
at its $B2$-$DO3$ disorder-order transition \cite{Bin91,HDBL80,Lei98}.
The values
$\eta_{\parallel}=1.52\pm0.04$, $\beta_1=0.75\pm0.06$, and
$\gamma_{11}=-0.33\pm0.06$ they found
are consistent with our estimates%
\footnote{The case of the $B2$-$DO3$ transition is more complicated
than that of the $A2$-$B2$ transition, for the $DO3$ structure involves
four sublattices and hence a larger number of
composition variables \cite{Bin91}. Nevertheless 
the $B2$-$DO3$ transition is expected to belong to the Ising universality
class \cite{HDBL80,Lei98}; see the note added in proof in
Ref.~\cite{Die97}.}
$\eta_{\parallel}^{\text{ord}}(n\!=\! 1)\simeq 1.53$,
$\beta_1^{\text{ord}}(n\!=\! 1)\simeq 0.80$,
and $\gamma_{11}^{\text{ord}}(n\!=\! 1)\simeq -0.33$
(taken from the last column of Table \ref{To1}).

An x-ray scattering experiment has also been performed
on the A2-B2 disorder-order transition in a semi-infinite FeCo alloy
that is bounded by a (001) surface \cite{KDN+97}. This yielded
$\beta_1=0.79\pm 0.10$, in conformity with the above theoretical values for $\beta_1^{\text{ord}}(n\!=\!1)$. Yet it should be noted that
the chosen (001) surface {\it breaks\/} the symmetry of
interchanging the two sublattices \cite{Sch92,DLBD97,LD98}.
Therefore the Hamiltonian one
encounters in a coarse-grained continuum description of the large-scale
physics is {\it not\/} invariant under a sign change
$\phi\to -\phi$ of the order
parameter and will generically have surface contributions
involving {\it odd\/} powers of $\phi$ and its
derivatives \cite{DC91,hwd}. In particular, surface contributions
{\it linear\/} in $\phi$, i.e., a {\it surface ordering field\/} $g_1\ne 0$, normally should be present, and since $g_1$ is a relevant scaling field, the
asymptotic critical behavior must be characteristic
of the {\it normal\/}%
\footnote{A justification of the name ``normal surface
transition'' may be found, for example, in Refs.\ \cite{Die94b}
and \cite{Die97}.}
\cite{Die94b,Die97} rather than the ordinary transition \cite{DLBD97,LD98}.

In their experiment, Krimmel {\it et al.}\ \cite{KDN+97} actually found
evidence of the presence of such a surface ordering field $g_1$.
On the other hand, they did not observe the crossover to the normal
surface transition. The reason seems to be that
$g_1$ is rather small. In order to
see clear manifestations of this crossover or even
verify the true asymptotic behavior,
one must therefore resolve a temperature
regime fairly close to $T_c$. The one
studied in the experiment was apparently not close
enough, a possibility which has already been suggested by the
experimentalists themselves \cite{KDN+97}.
A recent reanalysis \cite{Rit98} of their data
indicates that these are even better consistent
with the behavior one should expect near $T_c$ when
the scaling variable $g_1|\tau|^{-\Delta_1^{\text{ord}}}$ is still small
(so that the crossover to the normal surface transition has not yet set in)
than the original analysis by Krimmel {\it et al.} revealed.

The experiments \cite{dmjp} on the $B2$-$DO3$ transition of FeAl
also require a comment. Just as in the measurements on FeCo \cite{KDN+97},
a small amount of long-range order near the surface
was found to persist at and above $T_c$. It is tempting to attribute
this again to the presence of a surface ordering field $g_1$ (cf.\ Ref.\
\cite{Dos92}). However, the orientation of the surface plane
of the FeAl crystal investigated in Ref.\ \cite{dmjp} was
symmetry {\it preserving\/} in the sense of
Refs.\ \cite{Sch92} and \cite{LD98}, so surface contributions
breaking the $\phi\to -\phi$ symmetry of the Hamiltonian should not
occur. Thus, if the explanation of the
experimental findings must indeed be sought
in the presence of a surface ordering field, then the question
of its origin arises.%
\footnote{Any real surface will, of course, not
be ideally planar. Hence the symmetry invoked in proving the absence
of symmetry-breaking terms in the Hamiltonian \cite{LD98}
will not be strictly realized. Nevertheless, one would not expect
such unavoidable imperfections to manifest themselves
through symmetry-breaking contributions proportional to the surface area,
unless the crystal was not carefully prepared and its
surface not well-defined.}
It appears that further theoretical and experimental work 
is required to clarify this issue.

X-ray scattering experiments have also been
performed on a NH$_4$Br single crystal \cite{BPH93}.
The authors argue that the critical fluctuations
at the observed order-disorder transition should be described by
the three-dimensional Ising model, but also point
out that the transition is coupled to a first-order displacive transition.
The effective exponents they measured,
$\eta_\|=1.3\pm 0.15$ and $\beta_1=0.8\pm 0.1$,
are compatible with the theoretical predictions for the $n=1$ ordinary
transition. In view of the coupling to the displacive transition it is however
not clear to us how serious such a comparison can be taken.

Our estimates for  $n=2$ and $3$,
given in Tables \ref{To2} and \ref{To3},
also conform nicely with the previous $\epsilon$-expansion
estimates gathered in Table VI (p.~186) of Ref.\ \cite{hwd},
from which we quote the value $\eta_\|^{\text{ord}}(n\!=\!2)\simeq 1.38$
as an example (to be compared with our present best estimate
$\simeq1.42$). For $n=2$, there exist some recent 
Monte-Carlo results by Landau {\it et al.} \cite{lpb89},
as mentioned in Table \ref{To5}. For a comparison with
series-expansion estimates for the cases $n=2$ and $3$,
we refer to Table VII of Ref.\ \cite{hwd} and the original work \cite{OOM84}.

Our values $\eta_\|^{\text{ord}}(n\!=\!3)\simeq 1.34$ and
$\beta^{\text{ord}}_1(n\!=\!3)\simeq 0.82$ are fairly close to the estimates
$\eta_\|^{\text{ord}}(n\!=\!3)\simeq 1.29\pm 0.02$ and
$\beta^{\text{ord}}_1(n\!=\!3)\simeq 0.84\pm 0.01$
Diehl and N{\"u}sser \cite{DN86} obtained from Pad\'e approximants
that exploited the results of both the $\epsilon$ expansion
and the $d-2$ expansion to second order.
We are not aware of any recent
Monte-Carlo predictions for surface critical exponents of the $n=3$
ordinary transition.
On the experimental side, there is the
result  $\beta_1=0.825^{+0.025}_{-0.040}$
Alvarado {\it et al.}\ \cite{alvar} found for a Ni(100) surface
using spin-polarized low-energy electron diffraction.

\section{CONCLUDING REMARKS}\label{Concl}

In this work we have extended the massive field-theory approach
for studying critical behavior in a fixed space dimension
below the upper critical dimension  to systems with surfaces.
We have carried out two-loop calculations for the ordinary and special
surface transitions in $d=3$ bulk dimensions and performed
a Pad\'e-Borel analysis of the resulting series for the
respective surface critical exponents. The behavior of the
truncated series we have obtained and analyzed, though less good-natured
for some than for other exponents, is in general
very similar to what one finds for those of bulk exponents
at the same two-loop order of truncation. We take this as a clear indication
of the potential power of the approach: when pushed to an order of
perturbation theory that is comparable to what has been
achieved for the bulk exponents
\cite{ZJ,bgmn,bnm78,lgz80,AS95}
and investigated by the same sophisticated techniques based on
Borel summation and large-order analysis,
it should yield similarly precise numerical estimates.

One motivation for the present study was to see whether the
field-theory results might be reconciled with the
small values of $\simeq 0.5$ found
in recent Monte Carlo simulations \cite{rdww,hg94}
for the crossover exponents $\Phi(d\!=\! 3,n)$ with $n=0$ and $n=1$.
Our present best estimates
$\Phi (3,n\!=\!0)\simeq 0.52$ and $\Phi (3,n\!=\!1)\simeq 0.54$
(cf.\ Tables \ref{Tsp3} and \ref{Tsp4})
are indeed much lower than the original ones
based on the $\epsilon$ expansion (which were
$\simeq 0.67$ and $\simeq 0.68$, respectively \cite{dd83,hwd}),
and as we have seen, a Pad\'e-Borel analysis of the
$\epsilon$ expansion to order $\epsilon^2$
yields comparatively low $d=3$ estimates.
That the original $\epsilon$-expansion estimates
for $\Phi$ were $\simeq 20\%$ greater than our present ones
seems to be due to the unusual largeness of its
${\cal O}(\epsilon^2)$ terms, which entails that the value
of the truncated power series at $\epsilon =1$ gives a rather poor
approximation for $\Phi(d\!=\!3)$. This problem exist, of course,
also for the other surface exponents that derive from the same
RG function $\eta_{\sf c}$ as $\Phi$ (such as $\alpha_1^{\text{sp}}$,
cf.\ Tables \ref{Tsp3} and \ref{Tsp4}).
For the remaining surface exponents of both transitions,
the ${\cal O}(\epsilon^2)$
terms are much smaller, so the values of the truncated series at
$d=3$ turn out to be much closer to our best estimates.

In those cases in which there is little difference between the
$\epsilon$-expansion values given in Ref.\ \cite{hwd} and our
best estimates here, one may say that these field-theory values
have been put on a more reliable basis by our present analysis
because of our use of better extrapolation procedures based
on Pad\'e-Borel summation techniques.

To give error bars for our estimates of surface critical exponent
is a rather delicate matter. If we took as a measure of uncertainty
for the value of any given one of them
the spread of values of the various extrapolations of
the  ${\cal O}(u^2)$ series expansion, then a reasonable
guess might be a typical accuracy of a few, say, 5\%.
What appears to be needed most for an
improvement of the accuracy
and more reliable error bars is the computation of the
series coefficients of the surface exponents
to a higher order in perturbation theory.

There is an additional problem one is faced with in massive
field-theory approaches to systems with boundaries that
should be mentioned: the appearance of further mass scales
such as the renormalized surface enhancement $c$.
Having to deal with more than one mass parameter,
namely with $m$ and the ratio  $c/m$,
makes calculations rather cumbersome. Fortunately,
we have found ways to study directly the
asymptotic cases $c/m=0$ and $c/m\to\infty$ corresponding to
the  special and ordinary transitions, respectively.
Hence one gets back to single-mass problems.
Nevertheless the technical problems that must be overcome
to extend the calculations to higher orders of the loop expansion
require considerably more effort than in the bulk case.

It is our hope that the present work might serve as a useful basis and
starting point for further analyses that ultimately could lead to
quantitiative field-theory results for surface critical exponents
and other universal quantities of a precision as good as in the bulk
case. Finally, we would also like to express our hope that our work
might spur further experimental work as well as
simulations, the latter especially for higher spin dimensionalities.

\section*{Acknowledgements}

We are grateful to the Alexander von Humboldt (AvH) foundation
for awarding a research fellowship to one of us (M.S.). This
was vital for initiating the present work. Equally important
has been the support provided
by the Deutsche Forschungsgemeinschaft
(DFG) via Sonderforschungsbereich 237 and the Leibniz program
in subsequent stages of our work.

\appendix
\section{Two-loop calculations for the special transition}\label{asp}

In this appendix some details of our two-loop calculations
for the special transition will be presented.

\subsection{Mass renormalization}\label{amr}

Our starting point is the Feynman graph expansion
of $[G^{(0,2)}(p)]^{-1}$ to two-loop order,
as given by  (\ref{sigma0p}) and (\ref{sigma0p2l}).
A useful observation is that the term $C_3(p)$
in (\ref{sigma0p2l})
can be combined with the one $\propto [C_1(p)]^2$
as
\def\epsfsize#1#2{0.6#1}
\begin{equation}\label{C3D}
C_3(p)-{[C_1(p)]^2\over\kappa_0+c_0}=\;\;
\raisebox{-0.6em}{\epsfbox{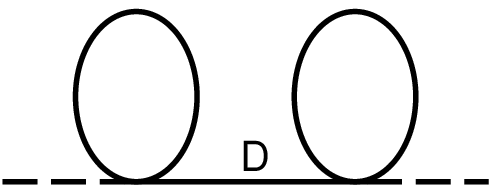}}\;\;\;,
\end{equation}
where the line labeled ``D'' represents the difference
\begin{equation}
\hat G(\bbox{p};z,z')-\frac{1}{c_0+\kappa_0}\,e^{-\kappa_0(z+z')}\;,
\end{equation}
which is nothing but the Dirichlet propagator
\begin{equation}\label{dirprop}
\hat G_{\text{D}}(\bbox{p};z,z')=
{1\over 2\kappa_0}\left[e^{-\kappa_0|z-z'|}-
e^{-\kappa_0(z+z')}\right].
\end{equation}

We now perform the mass renormalization. That is, we express
the bare mass $m_0$ in terms of  $m$ and $u_0$.
For our choice of bulk normalization conditions made in Sec.\ \ref{BNC},
one has (see, e.g., Ref.\ \cite{amit})
\begin{equation}
m_0^2=m^2-{n+2\over 6}\, u_0\, I_1(m^2)+{n+2\over 18}\,
 u_0^2\left[ I_2(m^2)
- m^2\,I_3(m^2)\right]+{\cal O}(u_0^3) \;,
\label{m0}\end{equation}
where 
\begin{equation}
I_1(m^2)=\int_{\bbox{k}}{1\over k^2+m^2}\;,
\label{j1}\end{equation}
\begin{equation}
I_2(m^2)\equiv I_2(0,m^2)\;,
\end{equation}
and
\begin{equation}
I_3(m^2)\equiv\left.{\partial\over\partial k^2}
I_2(k^2,m^2)\right|_{k^2=0}
\end{equation}
with
\begin{equation}
I_2(k^2,m^2)=\int_{\bbox{k}_1}\int_{\bbox{k}_2}
{1\over (k_1^2+m^2)(k_2^2+m^2)
\left[({\bbox{k}}_1+{\bbox{k}}_2+{\bbox{k}})^2+m^2)\right]}
\label{j2}\end{equation}
are familiar bulk Feynman integrals.
Unlike $I_1$ and $I_2$, the integral $I_3$
is finite in three dimensions,
\begin{equation}\label{I3}
I_3(m^2)=-{2\over 27}\left({1\over8\pi m}\right)^2\,.
\end{equation}
The factor $1/8\pi m$ may be identified as the value for $d=3$ of the
one-loop integral
\begin{equation}
D(m)=\int_{\bbox{k}}{1\over (k^2+m^2)^2}
\label{d2}\end{equation}
of the bulk vertex function
$\tilde\Gamma^{(4)}_{\text{bulk}}(\{\bbox{0}\})$.
In the massive field-theory approach, it is common
to absorb this factor in a properly defined
renormalized coupling constant. We did this when
introducing the rescaled renormalized coupling constant
$\tilde u$ via  (\ref{utildedef}). (Note that the rescaling factor
$b_n(d)$ in this equation becomes $[6/(n+8)]8\pi$ when $d=3$.)

So far (namely, in  (\ref{C3D}) as well as in the main text)
we used the convention that
the lines in Feynman graphs represent
the free and reduced propagators (\ref{freeprop})
and (\ref{redprop}), respectively.
For the purpose of determining the results
of the mass renormalization, it is preferable to
use a different one in which the full and dashed propagator
lines denote the modified expressions one obtains from the
previous ones through the replacement $m_0^2\to m^2$,
i.e., of $\kappa_0$ by $\kappa=\sqrt{p^2+m^2}$.

Let us decompose $\hat G$ into its bulk part
\begin{equation}\label{gb}
\hat G_{\text{b}}(\bbox{p};z,z')={1\over2\kappa}\,e^{-\kappa|z-z'|}
\end{equation}
and its surface part
\begin{equation}\label{gs}
\hat{G}_{\text{s}}(\bbox{p};z,z')={1\over2\kappa}{\kappa-c_0\over\kappa+c_0}
\ e^{-\kappa(z+z')}\;,
\end{equation}
writing
\begin{equation}\label{gdec}
\hat{G}(\bbox{p};z,z')=\hat G_{\text{b}}(\bbox{p};z,z')
+\hat G_{\text{s}}(\bbox{p};z,z')\;,
\end{equation}
and label lines representing $\hat{G}_{\text{b}}$ and $\hat{G}_{\text{s}}$ by 
``b'' and ``s'', respectively.
Since the term linear in $u_0$ in  (\ref{m0})
is given by the Feynman diagram
\begin{equation}
\epsfbox{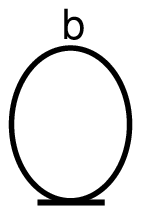}\;,
\end{equation}
it is clear that the perturbative corrections
originating from this term provide subtractions for the tadpole
subgraphs such that only their surface part
\begin{equation}\label{stp}
\epsfbox{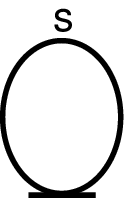}
\end{equation}
remains. Thus the self-energy
$\hat\sigma$ introduced in  (\ref{sigmap}) becomes
\begin{eqnarray}\label{sigmapr}
\hat{\sigma}(p;m,c_0)&=&
\;\raisebox{0.0em}{\epsfbox{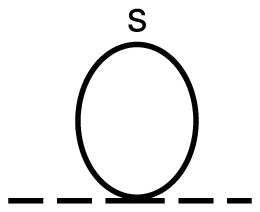}}\;+\hat\sigma_2(p;m,c_0)+\;\raisebox{0.0em}{\epsfbox{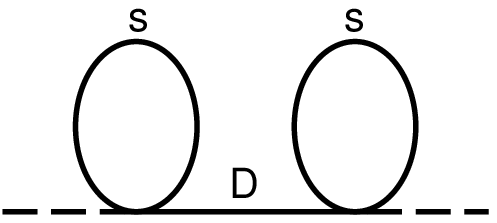}}\;
+\;\raisebox{0.0em}{\epsfbox{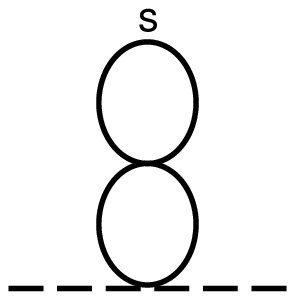}}\;+{\cal O}(u_0^3)
\end{eqnarray}
with
\begin{equation}\label{sigma2def}
\hat\sigma_2(p;m,c_0)=\;\vcenter{\epsfbox{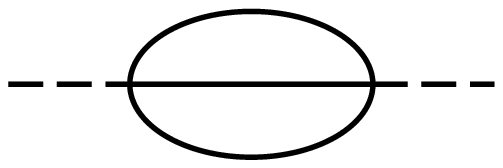}}\;\;
-\frac{1}{2\kappa}\,{n+2\over 18}\,
 u_0^2\left[ I_2(m^2)
- m^2\,I_3(m^2)\right].
\end{equation}
The latter combination, $\hat\sigma_2$, is uv finite for $d<4$.
However, the contributions of the  other three graphs
in (\ref{sigmapr}), which we shall denote as
 $\hat\sigma_1$, $\hat\sigma_3$,
and $\hat\sigma_4$, respectively, are not because they still
contain an uv divergence that must
be absorbed through the renormalization of
the surface enhancement.

\subsection{Surface-enhancement renormalization}\label{asr}

Upon substituting $\hat G^{(0,2)}$ from (\ref{sigmap})
into the normalization condition (\ref{cr}), one is led to
an equation for
the shift $\delta c$ introduced in (\ref{deltac}):
\begin{equation}\label{deltacit}
\delta c=\left\{\left[Z_1(c)Z_\phi\right]^{-1}-1\right\}
(c+m)+\hat{\sigma}(0;m,c_0=c+\delta c)\;.
\end{equation}
This can be solved by iteration to
determine the dependence of $c_0=c+\delta c$
on $c$ and $m$, order by order of perturbation
theory in $u_0$ (or $u$). Using it
(\ref{sigmap}) can be rewritten as
\begin{equation}\label{xxt}
\big[\hat G^{(0,2)}(p)\big]^{-1}=\kappa-m+
\left[Z_1(c)Z_\phi\right]^{-1}
(c+m) -\left[ \hat\sigma(p;m,c+\delta c)
-\hat\sigma(0;m,c+\delta c)\right]\;.
\end{equation}
Graphically the result can be interpreted as follows:
To obtain $\big[\hat G^{(0,2)}(p)\big]^{-1}$
one can set $\delta c=0$ if each graph of $\hat\sigma$
is taken to represent the graph itself subtracted by its value at $p=0$
and the corresponding subtractions are also made for subgraphs
involving lower-contributions to $\hat\sigma$. This is analogous to the familiar
graphical interpretation of mass renormalization in the bulk vertex function
$\tilde\Gamma^{(2)}(q)$ (see, e.g., \cite{Wallace}).

Let us be more specific in the special case $c=0$. Upon setting $c=0$
in (\ref{deltacit}), we see that the ${\cal O}(u_0)$ term of
the shift $\delta c$ can be written as
\begin{equation}\label{deltac1}
\delta c^{(1)}=[(Z_1^{\text{sp}}Z_\phi)^{-1}]^{(1)}\,m+
\hat\sigma_1(0;m,c_0\!=\!0)\;.
\end{equation}
The ${\cal O}(u_0)$ term  $[(Z_1^{\text{sp}}Z_\phi)^{-1}]^{(1)}$
will be calculated in the next subsection. The second contribution to $\delta c^{(1)}$
is uv singular for $d=3$. It provides the singular part of the counterterm
that cancels the uv divergence of $\hat\sigma_1(p;m,0)$. To show this,
let us calculate this quantity. We have
\begin{eqnarray}\label{sigma10}
\hat\sigma_1(p;m,0)&=&
-\frac{u_0}{2}\,\frac{n+2}{3}\int_0^\infty\!dz\,e^{-2\kappa z}\int_{\bbox{p}'}\frac{1}{2\kappa'}\,e^{-2\kappa' z}=-\frac{u_0}{8}\,\frac{n+2}{3}\int_{\bbox{p}'}
\frac{1}{(\kappa'+\kappa)\kappa'}\nonumber\\[1em]
&=&-\frac{u_0}{8}\,\frac{n+2}{3}\,m^{d-3}\,
K_{d-1}\,J\!\left({p\over m};d\right),
\end{eqnarray}
with $K_d=2^{1-d}\pi^{-d/2}/\Gamma(d/2)$, where
\begin{equation}
J(P;d)=\frac{1}{2} \int_0^\infty
\frac{y^{(d-3)/2}\,dy}{\left(\sqrt{P^2+1}+\sqrt{y+1}\right)\sqrt{y+1}}=
\int_0^\infty\!dy\,
y^{(d-3)/2}\frac{d}{dy}\ln\left[
\sqrt{P^2+1}+\sqrt{y+1}
\right].
\end{equation}
Subtracting from the latter integral its value at $P=0$ yields
the finite $d=3$ expression
\begin{eqnarray}\label{JP}
[J(P;d)-J(0;d)]_{d=3}&=&-\ln{{1+\sqrt{P^2+1}\over 2}}=
-\frac{1}{4} P^2+{\cal O}(P^4)\;.
\end{eqnarray}
These results can be substituted into the Dyson form (\ref{sigmap}) of
$\hat G^{(0,2)}$ to obtain
\begin{equation}
\left\{\hat G^{(0,2)}[p;m,c_0(c\!=\!0,m)]\right\}^{-1}=
\kappa+[(Z_1^{\text{sp}}Z_\phi)^{-1}]^{(1)}m-
\frac{n+2}{3}\;\frac{u_0}{8\pi}\,\ln{m+\kappa\over 2m}+{\cal O}(u_0^2)\;.
\end{equation}

At two-loop order, the first-order shift $\delta c^{(1)}$ must be
taken into account in $\hat\sigma_1$. It produces
an ${\cal O}(u_0^2)$ contribution of the form
\begin{equation}\label{onelct}
\delta c^{(1)}\,{\partial\hat\sigma_1\over\partial c_0}\,(p;m,c_0\!=\!0)\;,
\end{equation}
whose uv singular part must remove the divergence of the
tadpole subgraph (\ref{stp}) of $\hat\sigma_4$.
Since the same subgraph
appears (twice) in $\hat\sigma_3$ one may be surprised at the
apparent absence of the corresponding subtraction terms
one obtains through replacement of one or both of these subgraphs by
their singular parts%
\footnote{\label{sp}In the dimensionally regularized theory,
the singular part of (\ref{stp}) has the form
$(d-3)^{-1}\,\delta(z)$; if one rather employs a cutoff
regularization scheme, restricting the parallel
momentum integrations by $p\le\Lambda$, then
the pole gets replaced by $\ln\Lambda$.}%
, and one may wonder whether $\hat\sigma_3$ be uv finite.
In fact, the analog of $\hat\sigma_3$ that has the Dirichlet
line ``D'' replaced by a full propagator line would need such
subtractions, and those would occur, had we not
reorganized the expansion in the manner of (\ref{C3D})
to obtain graphs whose internal vertices $z$ and $z'$
are connected by the Dirichlet propagator
$\hat G_{\text{D}}(\bbox{p};z,z')$. The contribution $\hat\sigma_3$
is uv finite because $\hat G_{\text{D}}$ vanishes whenever $z$ or $z'$
approach zero. Replacing one or both tadpoles (\ref{stp}) in
$\hat\sigma_3$ by their singular parts\hbox{$^{\ref{sp}}$}
$\propto\delta(z)$ and $\propto \delta(z')$
would produce vanishing results.

According to (\ref{zsp}), we just need the expansion
coefficients of order $p^2$ of $[\hat G^{(0,2)}(p)]^{-1}$
to determine $Z_1^{\text{sp}}Z_\phi$. Therefore we shall
refrain from deriving the full two-loop result for
$[\hat G^{(0,2)}(p)]^{-1}$ here.
Before turning to the calculation of $Z_1^{\text{sp}}Z_\phi$,
let us compute the derivative $\partial\hat\sigma_1/\partial c_0$
in (\ref{onelct}). This is finite for $d=3$, and its ${\cal O}(p^2)$
coefficient will be needed
below. A straightforward calculation gives
\begin{eqnarray}\label{dsigmadc0}
{\partial\hat\sigma_1\over\partial c_0}\,(p;m,c_0\!=\!0)
&=&u_0\,\frac{n+2}{12}\,\int_{\bbox{p}'}\frac{1}{(\kappa'+\kappa)(\kappa')^2}=
\frac{u_0}{8\pi m}\,\frac{n+2}{3}\,
\frac{m}{m+\kappa}\ln {m+\kappa\over m}\\[1em]
 &=& \frac{u_0}{8\pi m}\,\frac{n+2}{3}\left[
\ln 2 + \frac14 \left(1-\ln4 \right)\frac{p^2}{m^2}
\right]+{\cal O}(p^4)\;.
\end{eqnarray}

\subsection{Calculation of $Z_1^{\text{sp}}Z_\phi$}

Using the representation (\ref{zsp}) of this renormalization factor,
one can easily deduce from  (\ref{sigma10}) and (\ref{JP})
its ${\cal O}(u_0)$ term:
\begin{equation}
\big[(Z_1^{\text{sp}}Z_\phi)^{-1}\big]^{(1)}=
-2m\left.\frac{\partial \hat\sigma_1(p;m,0)}{\partial p^2}\right|_{p=0}
=-{n+2\over 12}\,{u_0\over 8\pi m}\;.
\end{equation}
The terms of order $u_0^2$ can be written as
\begin{eqnarray}\label{Z1Zphistart}
\big[(Z_1^{\text{sp}}Z_\phi)^{-1}\big]^{(2)}&=&
\left({n+2\over 3}\,{u_0\over 8\pi m}\right)^2{1-\ln 4\over 8}
\nonumber\\&&\mbox{}
-2m\sum_{j=2}^4\,{\partial\over\partial p^2}\!\left[\hat\sigma_j(p;m,0)+\delta_{j4}\,\hat\sigma_1(0;m,0)\,
\frac{\partial\hat\sigma_1}{\partial c_0}(p;m,0)\right]_{p=0}.
\end{eqnarray}
The first term is the part of (\ref{onelct}) produced by the contribution
$[(Z_1^{\text{sp}}Z_\phi)^{-1}]^{(1)}$ of the shift (\ref{deltac1});
the remaining part of (\ref{onelct}) is the
last term in the square brackets.

To compute the contribution of $\hat\sigma_2$, we
start from
\def\epsfsize#1#2{0.5#1}
\begin{equation}\label{NNN}
\left.\vcenter{\epsfbox{G4m.eps}}\right|_{c_0=0}=\frac{n+2}{18}\,
u_0^2\int_0^\infty{\!dz}\,
\int_0^\infty\!dz'\,e^{-\kappa (z+z')}\int d^{d-1}r\left[G_{\text{N}}(\bbox{x},\bbox{x}')\right]^3\,
e^{i\bbox{p}\cdot \bbox{r}}\;,
\end{equation}
where $\bbox{x}=(\bbox{x}_\|,z)$, $\bbox{x}'=(\bbox{x}'_\|,z')$,
$\bbox{r}\equiv\bbox{x}_\|- \bbox{x}'_\|$,
and $G_{\text{N}}$ denotes the Neumann propagator. Introducing
\begin{equation}\label{radpm}
R_\pm=\sqrt{r^2+(z\pm z')^2}\;,
\end{equation}
we can write the latter as
\begin{equation}\label{ndecomp}
G_{\text{N}}(\bbox{x},\bbox{x}')=G_{\text{b}}(R_-)
+ G_{\text{b}}(R_+)\;.
\end{equation}
Since the term (\ref{NNN}) is not uv finite for $d=3$,
one must use the form
of the bulk propagator for general values of $d$,
\begin{equation}
G_{\text{b}}(x)=(2\pi )^{-{d\over 2}}\left(\frac{m}{x}\right)^{{d\over 2}-1}
K_{{d\over 2}-1}(mx)\;,
\end{equation}
rather than the simpler one for $d=3$,
\begin{equation}\label{gtryv}
G_{\text{b}}(x)=\frac{e^{-mx}}{4\pi\,x}\;.
\end{equation}

The derivative $\partial/\partial p^2|_0$
may then be performed under the integral sign. Recalling the definition
(\ref{sigma2def}) of $\hat\sigma_2$, one thus arrives at
\begin{equation}\label{sigma2cont}
-2m\left.\frac{\partial \hat\sigma_2(p;m,0)}{\partial p^2}\right|_{p=0}
=\frac{n+2}{18}\,u_0^2\left[\sum_{\lambda,\,\mu,\,\rho=\pm}
\left(I^{\lambda \mu\rho}+\frac{m}{d-1}\,J^{\lambda \mu\rho}\right)
-\frac{I_2(m^2)}{2m^2}+\frac{1}{2}\,I_3(m^2)\right],
\end{equation}
where we have introduced the integrals
\begin{equation}
I^{\lambda\mu\rho}=\int_0^\infty{\!dz}\,
\int_0^\infty\!dz'\,(z+z')\,e^{-m (z+z')}\int d^{d-1}r\,G_{\text{b}}(R_\lambda)\,
G_{\text{b}}(R_{\mu})\,G_{\text{b}}(R_\rho)
\end{equation}
and
\begin{equation}
J^{\lambda\mu\rho}=\int_0^\infty{\!dz}\,
\int_0^\infty\!dz'\,e^{-m (z+z')}\int\!d^{d-1}r\,r^2\,G_{\text{b}}(R_\lambda)\,
G_{\text{b}}(R_\mu)\,G_{\text{b}}(R_\rho)\;.
\end{equation}

Owing to the bulk singularity
contained in the distribution $[G_{\text{b}}(R_-)]^3$,
$I^{---}$ is not regular at $d=3$. Its singularity must be canceled
by that of the subtrahend $\propto I_2(m^2)$,
a term which can be cast in the form
\begin{equation}
\frac{I_2(m^2)}{2m^2}=\frac{1}{m^2}\int_0^\infty\!dz
\int\! d^{d-1}r\,\left[G_{\text{b}}(R)\right]^3
\end{equation}
with $R=\sqrt{r^2+z^2}$. The symmetric double integral
$\int_0^\infty\!dz\int_0^\infty\!dz'$ of $I^{---}$ reduces to a single one,
if we rewrite it as $\int_0^\infty\!dz_-\int_{z_-}^\infty\!dz_+$ with
$z_\pm=z\pm z'$ and integrate over $z_+$. This yields
\begin{equation}
I^{---}-{I_2(m^2)\over 2m^2}=\frac{1}{m^2}\int_0^\infty
\!dz\left[(1+mz)e^{-mz}-1\right]\int\!
d^{d-1}r 
\,\left[G_{\text{b}}(R)\right]^3.
\end{equation}
We may now set $d=3$ and perform the $r$ integration. The result is
\begin{equation}
I^{---}-{I_2(m^2)\over 2m^2}= \frac{1}{32 \pi^2m^2}\,i^{---}
\end{equation}
with
\begin{mathletters}\label{ijintdef}
\begin{equation}\label{i---def}
i^{---}=\int_0^\infty
dz(e^{-mz}+mz e^{-mz}-1)
\left[{e^{-3mz}\over z} -3m\text{E}_1(3mz)\right]\;,
\end{equation}
where $\text{E}_1(x)$ is the exponential integral function \cite{as}.
This can be evaluated with the help of the tables \cite{pbm1} and
\cite{pbm2}. One gets
\begin{equation}\label{a---}
i^{---}=2-7\ln{4\over 3}\,.
 \end{equation}

The remaining integrals in (\ref{sigma2cont}) are uv finite for $d=3$,
so one can set $d=3$ from the outset. A straightforward calculation gives
\begin{equation}
j^{---}\equiv 16\pi^2m^3\,J^{---}=
-{5\over 6}+3\ln{4\over3}\;,
\end{equation}
\begin{equation}
i^{+++}\equiv 32\pi^2m^2\,I^{+++}={7\over 4}-6\ln{4\over3}\;,
 \end{equation}
and
\begin{equation}
j^{+++}\equiv 16\pi^2m^3\,\,J^{+++}=
-{31\over 12}+9\ln{4\over3}\;.
 \end{equation}

The  integrals $I^{\mp\pm\pm}$ and $J^{\mp\pm\pm}$
require a little more effort.
Consider $I^{-++}$, for example. In the coordinates $r$,
$x=(z-z')/r$, and $y=(z+z')/r$, the $r$ integration
becomes trivial, giving
\begin{eqnarray}
i^{-++}\equiv 32 \pi^2m^2\,I^{-++}&=&m^2 \int_0^\infty\! dr\int_0^\infty \!dx
\int_0^x\!dy\,
{rx\over \kappa_x^2\kappa_{y}}\,
e^{-mr(x+2\kappa_x+\kappa_y)}\nonumber\\[1em]
&=&\int_0^\infty\! dx\int_0^x\!dy\,
{x\over\kappa_x^2 \kappa_{y}\, (x+2\kappa_x+\kappa_{y})^2}
\end{eqnarray}
with $\kappa_x\equiv\sqrt{1+x^2}$. In a similar manner one derives
\begin{equation}
i^{+--}\equiv 32 \pi^2m^2\,I^{+--}=\int_0^\infty\! dx
\int_0^x\!dy\, {x\over \kappa_x\kappa_y^2\,
(x+\kappa_x+2\kappa_y  )^2}\;,
\end{equation}
\begin{equation}
j^{-++}\equiv 16\pi^2m^3\,J^{-++}=
\int_0^\infty\! dx\int_0^x\!dy\,
{1\over \kappa_x^2\kappa_y\, (x+2\kappa_x+\kappa_y)^3}\;,
\end{equation}
and
\begin{equation}\label{j+--def}
j^{+--}\equiv 16\pi^2m^3\,J^{+--}=
\int_0^\infty\! dx
\int_0^x\!dy\, {1\over \kappa_x\kappa_y^2\,
(x+\kappa_x +2\kappa_y)^3}  \;.
 \end{equation}
\end{mathletters}\noindent

These integrals could be evaluated numerically in the present form.
However, by making the familiar Euler substitutions, one can reduce the
integrands to rational functions. Then the $y$ integrations become
standard and can be done exactly, so that one is left with
single integrations over $x$.  Finally, the latter can be transformed into
integrations over the bounded interval $[0,1]$, which is more
convenient for numerical calculations.
In Table \ref{TabAB} we have listed the numerical values of
the integrals $i^{\mp\pm\pm}$ and $j^{\mp\pm\pm}$ together
with those of the analytically known integrals
$i^{---}$, $j^{---}$, $i^{+++}$, and $j^{+++}$.

\begin{table}[hbtp]
\caption[truc]{\label{TabAB} Numerical values
of  the integrals  $i^{\lambda\mu\rho}$ and $j^{\lambda\mu\rho}$ for $d=3$.}
\begin{tabular}{l||d|d|d|d}
 $\lambda\mu\rho$   & $---$ & $+--$ & $-++$ & $+++$\\
\hline
$i^{\lambda\mu\rho}$ & -0.0137745 & 0.1324282 & 0.0420126 & 0.0239076\\
$j^{\lambda\mu\rho}$ & 0.0297129 & 0.0114128 & 0.0075298 & 0.0058053\\
\end{tabular}
\end{table}

Substituting the above results and the value (\ref{I3}) of $I_3$ into
(\ref{sigma2cont}) finally yields
\begin{mathletters}
\begin{equation}\label{sigma2fr}
-2m\left.\frac{\partial \hat\sigma_2(p;m,0)}{\partial p^2}\right|_{p=0}
={n+2\over 3}\,\left({u_0\over 8\pi m}\right)^2\,A
\end{equation}
with
\begin{eqnarray}
 A&=&\frac{1}{3}\sum_{\lambda,\,\mu,\,\rho =\pm}
\left(i^{\lambda\mu\rho}+j^{\lambda\mu\rho}\right)
-\frac{1}{162}\nonumber\\
&=&\mbox{}{17\over 162}-{1\over3}\ln{4\over3}
+i^{+--}+i^{-++}+j^{+--}+j^{-++}
\simeq 0.202428\;.
\end{eqnarray}
\end{mathletters}

Next, consider $\hat\sigma_3$. Noting that the integrand is a
symmetric function of $z$ and $z'$, we find
\begin{equation}
\hat\sigma_3(p;m,0)={1\over 4}\left({n+2\over3}{u_0\over 8\pi}\right)^2
{1\over\kappa}\int_0^\infty\!dz'\int_z^\infty\!dz\, {e^{-2m(z+ z')}\over
zz'}\,e^{-2\kappa z'}
\,(1-e^{-\kappa z})\;.
\end{equation}
The integrals one obtains upon performing the derivative
$\partial/\partial p^2\big|_0$ can be worked out utilizing the
mathematical tables \cite{pbm1,pbm2,devd}. This gives%
\footnote{Note that formula 3.1.3.2 on p.\ 567 of Ref.\ \cite{pbm1}
is incorrect; its corrected form is given as 3.2.1.2 on p.\ 617 of
Ref.\ \cite{pbm2}.}
\begin{equation}\label{sigma3fr}
-2m\left.\frac{\partial \hat\sigma_3(p;m,0)}{\partial p^2}\right|_{p=0}=
{1\over 4}\left({n+2\over3}\ {u_0\over 8\pi m}\right)^2
\left[\text{Li}_2\Big(-{1\over 2}\Big)+{\pi^2\over12}
+{1\over2}\ln{2\over3} \right] ,
\end{equation}
where $\text{Li}_2(x)$ is the polylogarithm of order 2
(see, e.g.,  Ref.\ \cite{devd}) with the value
$\text{Li}_2(-{1\over 2})= -0.4484142\ldots$.

It remains to compute the contribution of the combination
$\hat\sigma_4+\hat\sigma_1\,{\partial\hat\sigma_1/\partial c_0}$.
The second term implies that from
the term associated with the lower loop of $\hat\sigma_2$,
\begin{equation}
\frac12\,\int_0^\infty\frac{dp'\,p'}{2\pi}\,\hat G^2_{\text{N}}(\bbox{p}';z,z')\;,
\end{equation}
its value at $z'=0$, multiplied by $e^{-2\kappa' z'}$
(the factor produced by
the broken lines of $\hat\sigma_1$), gets subtracted.
This entails that the contribution we are concerned with
takes the form
\begin{equation}
-2m\left.\frac{\partial}{\partial p^2}\right|_{p=0}
\left[\hat\sigma_4+\hat\sigma_1\,
\frac{\partial\hat\sigma_1}{\partial c_0}\right]\!(p;m,0)
=
{1\over 4}\left({n+2\over3}\, {u_0\over 8\pi m}\right)^2g_4
\end{equation}
with
\begin{equation}
g_4=
\int_0^\infty\!dz\int_0^\infty\!dz'\,
\frac{2z\,e^{-2(z+z')}}{z'}
\left[F(z,z')-F(z,0)\,e^{-2z'}
\right],
\end{equation}
where $F$ denotes a sum of exponential integral functions \cite{as}:
\begin{equation}
F(z,z')=\text{E}_1(2|z-z'|)+\text{E}_1(2z+2z')+2\text{E}_1(|z-z'|+z+z')\;.
\end{equation}
The integrals contributing to $g_4$ can be evaluated analytically.
One obtains
\begin{equation}
g_4=-{\pi^2\over 12}+\ln^2 2+\frac12 \ln\frac{3}{2}-\text{Li}_2\Big(-\frac{1}{2}\Big).
\end{equation}

Combining the above results yields
\begin{equation}\label{Z1Zphiu0}
(Z_1^{\text{sp}}Z_\phi)^{-1}=1-{n+2\over 12}\,\frac{ u_0}{8\pi m}
+{n+2\over3}
\left[{n+2\over 12}
\left( {1\over2}-\ln2+\ln^2 2\right)+ A\right]
\left(\frac{ u_0}{8\pi m}\right)^2
+{\cal O}(u_0^3)\;,
\end{equation}
from which  (\ref{endz1}) follows upon substitution of
\begin{equation}\label{vr}
{u_0\over8\pi m}={6\over n+8}\left(\tilde{u}+\tilde{u}^2\right)+{\cal O}(\tilde{u}^3)\;.
\end{equation} 

\subsection{Calculation $Z_{\phi_s^2}^{\text{sp}}$}

We start from the representation (\ref{zc}) of this renormalization factor,
perform the mass renormalization, and use (\ref{sigmap}) to
express $[\hat{G}^{(0,2)}]^{-1}$
in terms of $\hat\sigma$, obtaining
\begin{equation}
(Z_{\phi_s^2}^{\text{sp}})^{-1}=Z_1^{\text{sp}}Z_\phi\left[1-
{\partial\hat\sigma(0;m,c_0)\over\partial c_0} \right]_{c=0}\,.
\label{azf2}\end{equation}
Upon carrying out the surface-enhancement renormalization, we arrive at
\begin{eqnarray}\label{Zphis2start}
Z_{\phi_s^2}^{\text{sp}}&=&(Z_1^{\text{sp}}Z_\phi)^{-1}
+{n+2\over3}\,\frac{u_0}{8\pi m}\,\ln 2+
\left({n+2\over3}\frac{u_0}{8\pi m}\right)^2\left( {1\over2}-
{3\over 4}\ln2+\ln^2 2\right)\nonumber\\
&&\mbox{}+ \left[\hat\sigma_1(0;m,0)\,
{\partial^2\hat\sigma_1(0;m,c_0)\over\partial c_0^2}+
\sum_{j=2}^4
{\partial \hat\sigma_j(0;m,c_0)\over\partial c_0}
\right]_{c_0=0}+{\cal O}(u_0^3)
\end{eqnarray}

The explicitly given first ${\cal O}(u_0^2)$ contribution
is the sum of the three terms 
$(\partial\hat\sigma_1/\partial c_0)^2$,
$-(Z_1^{\text{sp}}Z_\phi)^{(1)}\partial\hat\sigma_1/\partial c_0$,
and $-(Z_1^{\text{sp}}Z_\phi)^{(1)}m\,\partial^2\hat\sigma_1/\partial c_0^2$,
whose last one is one part of the contribution proportional
to the ${\cal O}(u_0)$ shift $\delta c^{(1)}$ (the other part being
the contribution proportional to the uv singular tadpole graph $\hat\sigma_1$).
The values of $(Z_1^{\text{sp}}Z_\phi)^{(1)}$ and
$\partial\hat\sigma_1/\partial c_0$ may be read off from
(\ref{Z1Zphiu0}) and (\ref{dsigmadc0}), respectively;
the required second derivative is
\begin{equation}
{\partial^2\hat\sigma_1\over\partial c_0^2}\,(0;m,c_0\!=\!0)
=-u_0\,\frac{n+2}{6}\,\int_{\bbox{p}'}\frac{1}{(\kappa'+m)(\kappa')^3}=
-\frac{n+2}{3}\,\frac{u_0}{4\pi m^2}\left(1-\ln 2\right)\;.
 \end{equation}

The remaining contributions in (\ref{Zphis2start}) can be
computed by proceeding along lines similar to those taken
in our calculation of $Z^{\text{sp}}_1Z_\phi$.
The contribution of $\hat\sigma_2$ is given by the derivative $\partial/\partial c_0$ of the graph (\ref{NNN}) at $c_0=0$.
It can be written as
\begin{mathletters}
\begin{eqnarray}\label{sig2c_0}
\frac{\partial\hat\sigma_2}{\partial c_0}\,(0;m,c_0\!=\!0)
&=&-{n+2\over3}\left(\frac{u_0}{8\pi m}\right)^2\,B
 \end{eqnarray}
with
\begin{equation}\label{syin}
B=-{(8\pi m)^2\over 6}\,
\int_0^\infty{\!dz}\,
\int_0^\infty\!dz'\,e^{-m (z+z')}
\int d^{d-1}r \left.
\frac{\partial}{\partial c_0}\left[
G(\bbox{x},\bbox{x}')\right]^3 \right|_{c_0=0},
\end{equation}
\end{mathletters}\noindent
where again $\bbox{r}=\bbox{x}_\|-\bbox{x}'_\|$. The required
derivative
$(\partial G^3/\partial c_0)_{c_0=0}=
3G_{\text{N}}^2\,(\partial G/\partial c_0)_{c_0=0}$ can be computed
in the $\bbox{p}z$ representation. Upon setting $d=3$ and
performing the angular integration, we find
\begin{equation}\label{B2}
\left.\frac{\partial}{\partial c_0}\> G(\bbox{x},\bbox{x}')\right|_{c_0=0}=
-\int_{\bbox{p}} {e^{-\kappa(z+z')}\over\kappa^2}\,e^{-i\bbox{p}\cdot \bbox{r}}=-\int_0^\infty\!{dp\,p\over 2\pi\kappa^2}\,J_0(pr)\,e^{-\kappa(z+z')}\;.
\end{equation}
We insert this, together with the decomposition (\ref{ndecomp})
of $G_{\text{N}}$, into (\ref{B2}) and perform the angular part
of the integration $\int d^2r$. This yields
\begin{equation}
B=2\sum_{\lambda,\,\mu=\mp}k^{\lambda\mu}
\end{equation}
with
\begin{equation}
k^{\lambda\mu}= m^2
\int_0^\infty{\!dr}\,r\,
\int_0^\infty\frac{dp\,p}{\kappa^2}\,J_0(pr)
\int_0^\infty{\!dz}\,
\int_0^\infty\!dz' e^{-(\kappa+m) (z+z')}\,
\frac{e^{-m (R_\lambda+R_\mu)}}{R_\lambda R_\mu}\;,
\end{equation}
where $R_{\lambda=\mp}$ is defined by (\ref{radpm}).

Changing again to the variables $z_\pm=z\pm z'$, we can rewrite
the symmetric double integrals $\int_0^\infty dz\int_0^\infty dz'$
in $k^{--}$ and $k^{++}$ as
$\int_0^\infty dz_-\int_{z_-}^\infty dz_+$
and
$\int_0^\infty dz_+\int_0^{z_+}dz_-$, respectively. The
inner one of these $z$ integrations can now be performed to get
\begin{equation}\label{klamlam}
k^{\lambda\lambda}=
m^2\int_0^\infty{\!dr}\,r\,
\int_0^\infty\frac{dp\,p}{\kappa^2}\,J_0(pr)
\int_0^\infty{\!dz}\,
\frac{e^{-(\kappa+m)z-2mR}}{R^2}\times\cases{ (\kappa+m)^{-1}& for $\lambda=-\;,$\cr
z&for $\lambda=+\;,$\cr}
\end{equation}
where $R=\sqrt{r^2+z^2}$, as before.
Next we make the variable transformation
$r \to r/z$. Then the integrals over $z$ simplify to
\begin{equation}\label{simplifform}
\int_0^\infty{\!dz}\,
e^{-(\kappa+m+2mR) z} J_0(prz)\times\cases{1&for $\lambda=-\;,$\cr
z\;,&for $\lambda=+\;,$\cr}
\end{equation}
standard integrals, which can be found in tables \cite{pbm2}.
In this manner one obtains
\begin{mathletters}
\begin{equation}
k^{--}=
\int_0^\infty \frac{dr\,r}{\kappa_r^2}\,
\int_0^\infty\!\frac{dp\,p}{\kappa_p^2 \>(\kappa_p +1)
\sqrt{(1+\kappa_p+2 \kappa_r)^2 +(pr)^2}}
\end{equation}
and
\begin{equation}
k^{++}=
\int_0^\infty \frac{dr\,r}{\kappa_r^2}\,
\int_0^\infty\frac{dp\,p}{\kappa_p^2}\,
{1+\kappa_p+2 \kappa_r\over
\left[\big(1+\kappa_p+2 \kappa_r\big)^2 +(pr)^2 \right]^{3/2}}\;,
\end{equation}
\end{mathletters}\noindent
where as before $\kappa_p\equiv \sqrt{p^2+1}$.
To compute these integrals, we had to resort once more
to numerical means. Our results read
\begin{eqnarray}\label{numvalk}
k^{--}&=&0.1538951\,,\\ k^{++}&=&0.0469337\,.
\end{eqnarray}

The ``mixed'' term $k^{-+}$ is more complicated since
its double $z$ integration does not reduce
to a single one. We first rescale $z$, $z'$, and $r$, so that
$z/r \to z$, $z'/r \to z'$, and $mr\to r$. This leads us to
\begin{equation}
k^{-+}=
\int_0^\infty{\!dr}\,r
\int_0^\infty\frac{dp\,p}{\kappa_p^2}\,J_0(pr)
\int_0^\infty{\!dz}\,
\int_0^\infty\!dz'\,e^{-r\,(\kappa_p+1) (z+z')}
{e^{-r\sqrt{|z-z'|^2+1}}\over \sqrt{|z-z'|^2+1}}\,
{e^{-r \sqrt{(z+z')^2+1}}\over\sqrt{(z+z')^2+1}}\,.
\end{equation}
The integral over $r$ is of the form (\ref{simplifform}) with $\lambda=+$
and hence can be calculated. Rewriting $\int_0^\infty dz\int_0^\infty dz'$
as $\int_0^\infty dy\int_0^y dx$ with $y=z+z'$ and $x=z-z'$ finally gives
the representation we used to determine $k^{-+}$ by numerical integration:
\begin{equation}
k^{-+}=
\int_0^\infty\frac{dp\,p}{\kappa_p^2}
\int_0^\infty \frac{dy}{\kappa_y}
\int_0^y\frac{dx}{\kappa_x}\,
{ \kappa_y+\kappa_x+(\kappa_p+1)y 
\over\left\{ \left[ \kappa_y+\kappa_x+(\kappa_p+1)y \right]^2
+p^2 \right\}^{3/2}}\;.
\end{equation}
Our result for $k^{-+}$ is
\begin{equation}
k^{-+} \simeq 0.0691008\,.
\end{equation}
In conjunction with (\ref{numvalk}) it yields
\begin{equation}
B=2\left(k^{--}+2k^{-+}+k^{++}\right)\simeq 0.678061\,,
\end{equation}
which is the numerical value quoted in (\ref{Bnumval}).

The other two-loop contributions to
$Z_{\phi_s^2}^{\text{sp}}$ can be worked out analytically.
Performing first
the parallel-momentum loop integrations and thereafter
the remaining ones over $z$ and $z'$, one
finds that%
\footnote{Again, the tables of integrals \cite{pbm2}
were used; note that formula 2.5.11.11 is misprinted:
the argument of the exponential
function in the integrand should
read $(c-b)x$ instead of $(b-c)x$.}
\begin{equation}
\frac{\partial\hat\sigma_3}{\partial 
c_0}\,(0;m,0)=\left({n+2\over3}\,\frac{u_0}{8\pi 
m}\right)^2\left[\text{Li}_2\!\left(-{1\over2}\right)+{\pi^2\over12}+{1\over2}\ln2-{3\over4}\ln3\right]
\end{equation}
and
\begin{eqnarray}\label{sigm4c0}
\left(
\frac{\partial\hat\sigma_4}{\partial c_0}
+\hat\sigma_1{\partial^2\hat\sigma_1\over\partial c_0^2}
\right)(0;m,0)&=&\left({n+2\over3}\,\frac{u_0}{8\pi m}\right)^2
\left\{\left[ {\pi^2\over24}-{5\over2}\ln2+{3\over4}\ln3\right]\right.\nonumber \\[1em]
&&\mbox{}+\left.\left[
-\text{Li}_2\!\left(-{1\over2}\right)
-{\pi^2\over8}+\ln2+{1\over4}\ln^2 2\right]\right\}\nonumber\\[1em]
&=&\mbox{}
\left({n+2\over3}\,\frac{u_0}{8\pi m}\right)^2
\left[
\frac{3}{4}\,\ln\frac{3}{4}+\frac{1}{4}\ln^2 2-
\text{Li}_2\!\left(-{1\over2}\right)-{\pi^2\over 12}
\right].
\end{eqnarray}

In the first version of (\ref{sigm4c0}),
the term in the first pair of square brackets represents
the part of $\partial \hat\sigma_4/\partial c_0$ that results from
the action of the derivative $\partial/\partial c_0$
on the tadpole subgraph (\ref{stp}) of $\hat\sigma_4$
(by which an insertion of $\phi_s^2/2$ is produced there).
Whereas this part is uv finite at $d=3$, the other one
(with an insertion of $\phi_s^2/2$ at the lower loop)
must be combined with the remaining term $\propto \hat\sigma_1$ to
obtain the finite result pertaining to the second pair of
square brackets.

Combining these results yields
\begin{eqnarray}\label{Zphis2u0}
Z_{\phi_s^2}^{\text{sp}}&=&1+{n+2\over3}\,\frac{u_0}{8\pi m}
\left(\ln 2-\frac{1}{4}\right)\nonumber\\&&\mbox{}
+{n+2\over3}\left(\frac{u_0}{8\pi m}\right)^2\left[A-B
+{n+2\over3}
\left({1\over 4}\ln^2 2-\ln2\right)\right]
+{\cal O}(u_0^3)\;,
\end{eqnarray}
from which (\ref{endzc}) follows by expressing $u_0$ in terms
of $\tilde u$ using (\ref{vr}).
Just as in the case of $Z^{\text{sp}}_1Z_\phi$, the number
Li$_2(-1/2)$ has canceled out.

\section{Two-loop Feynman diagrams for the ordinary transition}\label{appord}

We first show that the last two two-loop contributions in (\ref{sigmapr}),
$\hat\sigma_3(p;m,c_0)$ and $\hat\sigma_4(p;m,c_0)$, sum to zero when
$c_0=\infty$:
\def\epsfsize#1#2{0.5#1}
\begin{equation}
\hat\sigma_3(p;m,\infty)+\hat\sigma_4(p;m,\infty)=
\left[\vcenter{\epsfbox{G3m.eps}}\;
+\;\raisebox{-1.4em}{\epsfbox{G2m.eps}}\;\right]_{c_0=\infty}=0\;.
\end{equation}
A straightforward calculation gives
\begin{equation}\label{sigmaD34}
\hat\sigma_{j}(p;m,\infty)=\frac{u_0^2}{128}\left(
{n+2\over 3}\right)^2\int_{\bbox{p}_1}\int_{\bbox{p}_2}\,
\frac{1}{\kappa_1\kappa_2}\,
L_j(\kappa,\kappa_1,\kappa_2)\;,\qquad j=3,4,
\end{equation}
with
\begin{eqnarray}
L_3(\kappa,\kappa_1,\kappa_2)&=&{4\over\kappa}\,
 \int_0^\infty\!dz\int_0^\infty\!dz'\,
e^{-\kappa(z+z')-2(\kappa_1z+\kappa_2 z')}\;
\left[e^{-\kappa|z-z'|}-e^{-\kappa(z+z')}\right]
\nonumber\\
&=&{1\over \kappa}\, {1 \over\kappa+\kappa_1+
\kappa_2} \left({1\over \kappa+\kappa_1}
+{1\over\kappa +\kappa_2}\right)
-{1\over \kappa}\, {1\over\kappa+\kappa_1}
{1\over\kappa+\kappa_2} \;.
\end{eqnarray}
and
\begin{eqnarray}
L_4(\kappa,\kappa_1,\kappa_2)&=&-{4\over\kappa_1}\,
 \int_0^\infty\!dz\int_0^\infty\!dz'\,e^{-2(\kappa z+\kappa_2 z')}\;
\left(e^{-\kappa_1|z-z'|}-e^{-\kappa_1(z+z')}\right)^2
\nonumber\\
&=&{1\over \kappa_1}\left[\left( {-1 \over\kappa+\kappa_2} 
+{2 \over\kappa+\kappa_1+\kappa_2}\right)
\left({1\over \kappa+\kappa_1}
+{1\over\kappa_1 +\kappa_2}\right)
- {1\over\kappa+\kappa_1}
{1\over\kappa+\kappa_2}\right].
\end{eqnarray}
Adding these contributions yields
\begin{equation}\label{L34}
\sum_{j=3}^4L_j(\kappa,\kappa_1,\kappa_2)=
{\kappa_1-\kappa_2 \over (\kappa+\kappa_1)(\kappa+\kappa_2)
(\kappa+\kappa_1+ \kappa_2)(\kappa_1+ \kappa_2)}\;.
\end{equation}
Hence the integral $\int_{\bbox{p}_1}\int_{\bbox{p}_2}$
in (\ref{sigmaD34}) is identically zero because of the antisymmetric
form of its integrand.

Therefore the ${\cal O}(u_0^2)$ term of $Z_{1,\infty}Z_\phi$
is indeed entirely given by the contribution of the quantity
$\hat\sigma_2(p;m,\infty)$ defined in (\ref{sigma2def}), as we
asserted in (\ref{z2l}). To compute it, we can
proceed as in our calculation of its $c_0=0$ analog
in Appendix \ref{asp}.
This leads us to
\begin{equation}
-2m\left.
\frac{\partial \hat\sigma_2(p;m,\infty)}{\partial p^2}
\right|_{p=0}=\frac{n+2}{3}\,\left({u_0\over 8\pi m}\right)^2 C
\end{equation}
with
\begin{equation}
C:=\frac{1}{3}\left(
i^{---}+j^{---}-i^{+++}-j^{+++}\right)-\frac{1}{162}
+i^{-++}+j^{-++}-i^{+--}-j^{+--}\;.
\end{equation}
Here $i^{---},\ldots,j^{+--}$ are the integrals introduced
in (\ref{i---def})--(\ref{j+--def}).
Using the analytical values given
in these equations together with the numerical ones
listed in Table \ref{TabAB}, one arrives at the form
(\ref{allzc0}) of our results for $Z_{1,\infty}Z_\phi$
and $C$.

\end{document}